\documentclass[aps,amsfonts,pra,twocolumn,showpacs]{revtex4-1}

\usepackage{subfigure}
\usepackage{textcomp}
\usepackage{graphicx}
\usepackage{amssymb}
\usepackage{amsmath}
\usepackage{bm}
\usepackage{amsmath, amsthm, amssymb}
\usepackage{dsfont}
\usepackage{url}

\newcommand{\mean}[1]{\left[ #1 \right]}

\newcommand{\cexp}[1]{\left< #1 \right>}

\newcommand{\std}[1]{\sigma{\left(#1\right)}}

\newcommand{\x}{\tau^x}

\newcommand{\z}{\tau^z}

\newcommand{\red}[1]{#1}

\def\s{s}

\def\beq{\begin{equation}}
\def\eeq{\end{equation}}
\def\bea{\begin{eqnarray}}
\def\eea{\end{eqnarray}}

\usepackage[usenames, dvipsnames]{color}

\begin{document}

\title{Quantum criticality in Ising chains with random hyperuniform couplings}

\author{P. J. D. Crowley$^1$, C. R. Laumann$^1$, S. Gopalakrishnan$^2$}
\affiliation{$^1$ Physics Department, Boston University, Boston, MA 02215, USA \\ $^2$ Department of Physics and Astronomy, CUNY College of Staten Island, Staten Island, NY 10314; Physics Program and Initiative for the Theoretical Sciences, The Graduate Center, CUNY, New York, NY 10016, USA}

\begin{abstract}
We study quantum phase transitions in transverse-field Ising spin chains in which the couplings are random but hyperuniform, in the sense that their large-scale fluctuations are suppressed. We construct a one-parameter family of disorder models in which long-wavelength fluctuations are increasingly suppressed as a parameter $\alpha$ is tuned. For $\alpha = 0$, one recovers the familiar infinite-randomness critical point. For $0 < \alpha < 1$, we find a line of infinite-randomness critical points with continuously varying critical exponents; however, the Griffiths phases that flank the critical point at $\alpha = 0$ are absent at any $\alpha > 0$. When $\alpha > 1$, randomness is a dangerously irrelevant perturbation at the clean Ising critical point, leading to a state we call the critical Ising insulator. In this state, thermodynamics and equilibrium correlation functions behave as in the clean system. However, all finite-energy excitations are localized, thermal transport vanishes, and autocorrelation functions remain finite in the long-time limit. We characterize this line of hyperuniform critical points using a combination of perturbation theory, renormalization-group methods, and exact diagonalization. 

\end{abstract}

\maketitle

\section{Introduction}

Quenched randomness has profound effects on the thermodynamics and dynamics of quantum systems. Equilibrium quantum critical points are unstable to weak randomness if the correlation length exponent violates the Harris criterion $\nu > 2/d$, where $d$ is the spatial dimension~\cite{Harris:1974aa}. When the clean critical point is unstable, the system might exhibit a random critical point~\cite{ccfs}, at which its properties are heterogeneous on all length-scales, or the phase transition might be rounded~\cite{vojta_rounding} or preempted---for instance, rare-region effects might destabilize one of the phases (for recent examples see Refs.~\cite{drh_stability, pixley_rare}). In addition, disorder qualitatively modifies quantum dynamics through Anderson localization of elementary excitations~\cite{lr1985}. In low-dimensional systems, the Harris bounds are particularly stringent, and besides, weak randomness localizes all excitations~\cite{lr1985}. Thus, at clean low-dimensional quantum critical points, both equilibrium properties and dynamics tend to be unstable to weak randomness. In the paradigmatic instance of the transverse-field Ising chain~\cite{sml1964}, the clean correlation length exponent $\nu = 1$, which violates the Harris criterion; for any disorder, the true critical point is at infinite randomness~\cite{fisher1992random, fisher1995critical, fisher1999phase}, and all excitations at nonzero energy are exponentially localized.

\begin{figure}[b]
\begin{center}
\includegraphics[width = 0.45\textwidth]{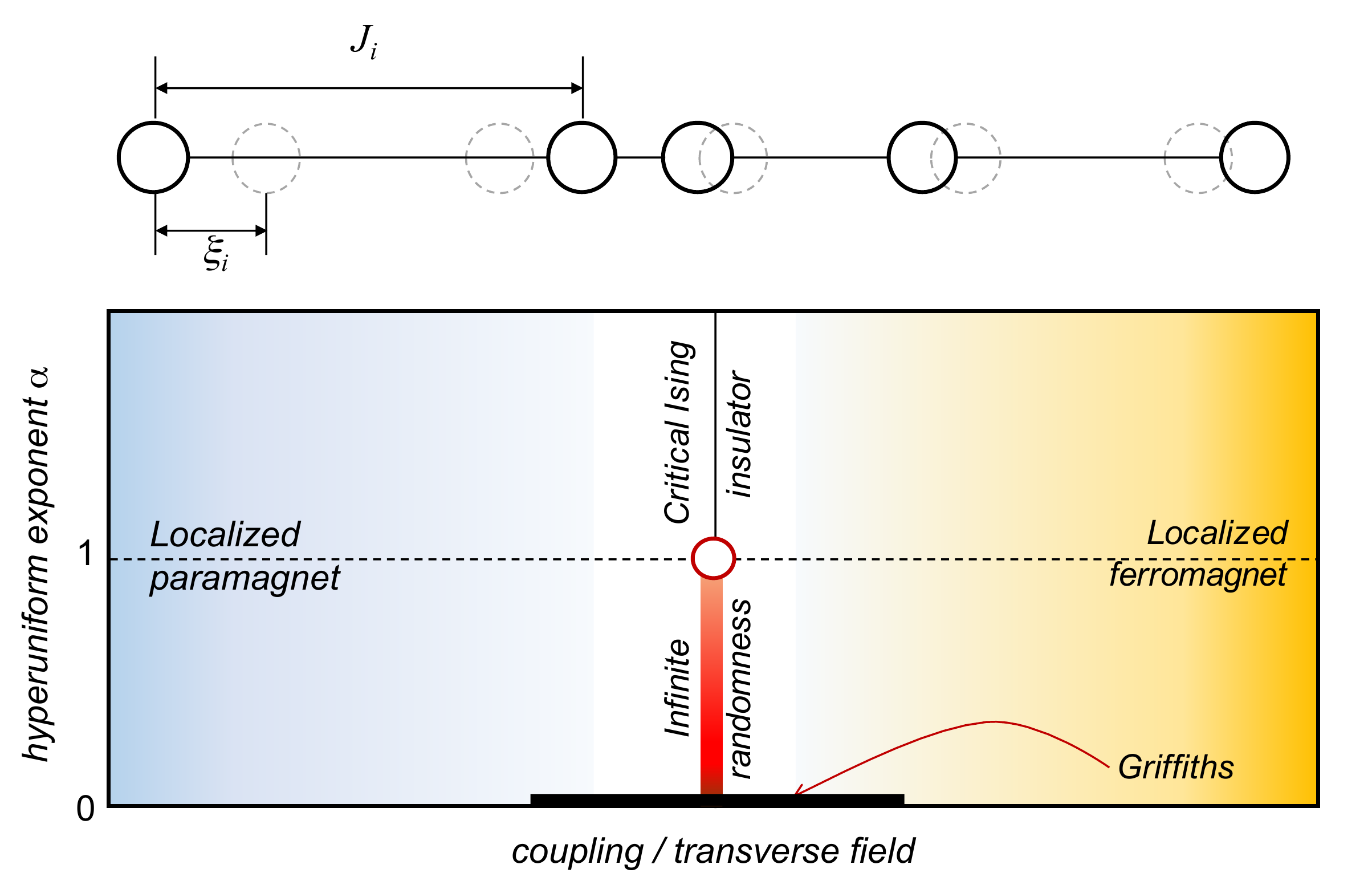}
\caption{Upper panel: simple model with strongly hyperuniform ($\alpha = 2$) bond randomness. Each \emph{site} is displaced by an independent random amount $\xi_i$ from its equilibrium position; this leads to bonds $J_i$ that are evidently strongly hyperuniform, in the sense that the variance of $\sum_{i = 1}^{l} J_i$ tends to a constant independent of $l$. Lower panel: Phase diagram of the random hyperuniform TFIM. Away from criticality, all excitations are localized for any randomness; however, the universality class of the critical point changes as the hyperuniformity parameter $\alpha$ is varied. A line of infinite-randomness critical points, with continuously varying exponents, terminates at a multicritical point, beyond which disorder is dangerously irrelevant (the critical Ising insulator).} 
\label{fig0}
\end{center}
\end{figure}

This standard analysis applies when the disorder lacks large-scale spatial correlations. However, localization and the instability of clean critical points occur more generally, even for deterministic quasiperiodic potentials~\cite{Azbel:1979aa, Aubry:1980aa, biddle2009localization, ganeshan2015nearest, lueschen_qp,  gopalakrishnan2017self, pwhg,  schreiber2015observation, iorh, ksh, lev2017transport, znidaric2018interaction, wgk}. Quasiperiodic couplings, when weak, neither affect critical properties nor localize excitations; thus, unlike random couplings, they are perturbatively irrelevant for both statics and dynamics. At a critical strength of the quasiperiodic potential, however, all excitations localize and the equilibrium critical point is concomitantly destabilized~\cite{igloi1988quantum, Chandran:2017ab, ccl, crowley2018critical}. That the onset of localization and the critical-point instability coincide in both random and quasiperiodic systems might suggest that they are somehow fundamentally linked; this is consistent with the intuition~\cite{sachdev_book} that statics and dynamics are inherently linked at quantum phase transitions. Conceptually, however, localization and the instability of critical points stem from different aspects of disorder: the former is due to the disorder potential having a continuous momentum-space spectrum; the latter, to long-wavelength fluctuations. Uncorrelated randomness has both features, while quasiperiodic potentials have neither. However, a broad class of \emph{random} patterns also have suppressed large-scale fluctuations. 
These patterns are called ``hyperuniform''~\cite{torquato2003, torquato2016, ma2017, torquato2018hyperuniform}. The local value of the order parameter $\delta_i$ at site $i$ in disordered system is characterised by its spatial average $\delta = [\delta_i]$ and with fluctuation scale set by the standard deviation $\sigma(\delta_i)$. Consider the integrated value $S_l(i) = \sum_{|j|< l} \delta_{i+j}$ summed over the region of linear size $l$ in $d$ dimensions and centred at site $i$. is characterised by an asymptotic mean $[S_l] \sim l^d \delta$. $S_l(i)$ For short range correlated $\delta_i$, the fluctuations $\sigma(S_l)$ are Poissonian, scaling as $\sigma(S_l) \sim l^{d/2}$. The spatial variation of $\delta_i$ is said to be hyperuniform if the fluctuations scale as $\sigma(S_l) \sim l^{\beta}$ with $\beta < d/2$. Even in maximally uniform structures, such as crystals, where $\delta_i$ has the periodicity of the lattice, the are still fluctuations of order $\sigma(S_l) \sim l^{(d-1)/2}$, in a period structure this contribution comes from the boundary. We shall refer to systems with $\beta = (d-1)/2$ as \emph{strongly hyperuniform} (Class I of Ref.~\cite{torquato2018hyperuniform}). Systems with intermediate exponent $(d - 1)/2 < \beta < d/2$, are \emph{weakly hyperuniform} (Classes II and III of Ref.~\cite{torquato2018hyperuniform}). 

This work considers quantum phase transitions in which the control parameter is spatially varying and exhibits random hyperuniform fluctuations. Previous analyses have extensively explored the implications of hyperuniform fluctuations in particle density, both theoretically and experimentally in photonic materials~\cite{florescu2009designer, wiersma2013disordered}, their localization properties have been studied numerically~\cite{froufe2017}; however, phase transitions in systems with hyperuniform couplings have not previously been explored. 
We focus on the transverse-field Ising model, subject to random hyperuniform bonds and/or transverse fields, with tunable extent of hyperuniformity; the picture that emerges from our study is quite general, however, and applies to a range of phase transitions in systems with hyperuniform couplings. 
The models we introduce are constructed in momentum space, and are thus simple to implement in ultracold atomic gases using spatial light modulators~\cite{zoran_slm} or in ``quantum gas microscope'' experiments~\cite{simon2011}. The hyperuniform couplings in the models we consider lead to a modification to the usual Harris criterion, which is obtained for uncorrelated spatial disorder. This alters the condition for the relevance of the disorder to the critical properties of the transition. Despite this change, as we consider one-dimensional models with local hopping subject to random bonds with continuous Fourier spectra, the $E \neq 0$ excitations remain subject to weak localization. Thus, these models are intermediate between random and quasiperiodic systems: all excitations localize at weak disorder, but the clean critical points need not be unstable. Hyperuniform disorder comes about due to correlations. The effect of correlated disorder on localization~\cite{planderson1, planderson2, planderson3} and phase transitions~\cite{weinrib, igloi1999, hoyos2011, chatelain2014} have been previously explored; these works, however, were concerned with the case of locally \emph{correlated} disorder, whereas the present work addresses local \emph{anticorrelations}, which naturally give rise to entirely different physics.

\begin{figure}[tb]
\begin{center}
\begin{minipage}{0.23\textwidth}
\includegraphics[width = 0.95\textwidth]{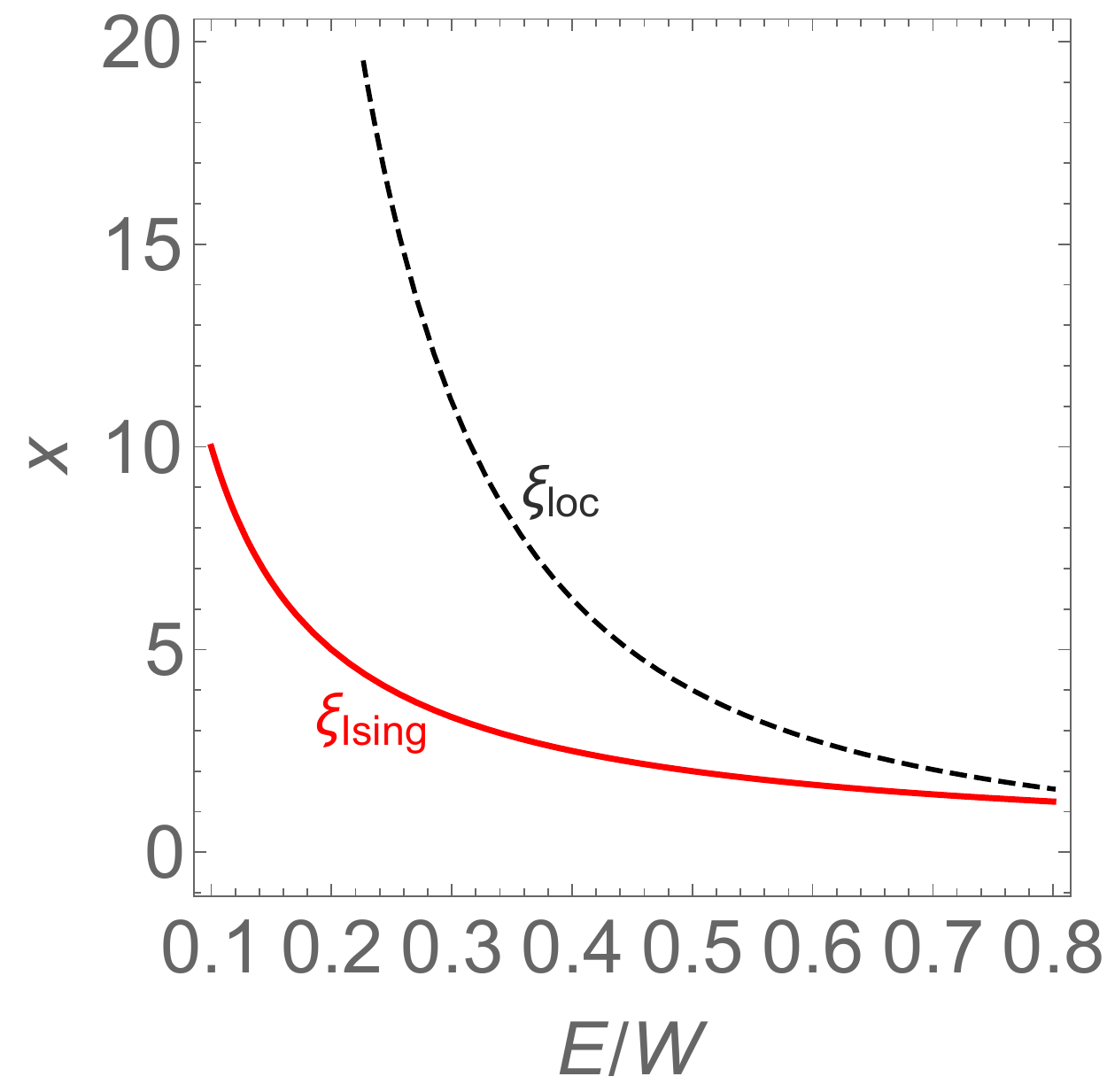}
\end{minipage}
\begin{minipage}{0.23\textwidth}
\includegraphics[width = 0.95\linewidth]{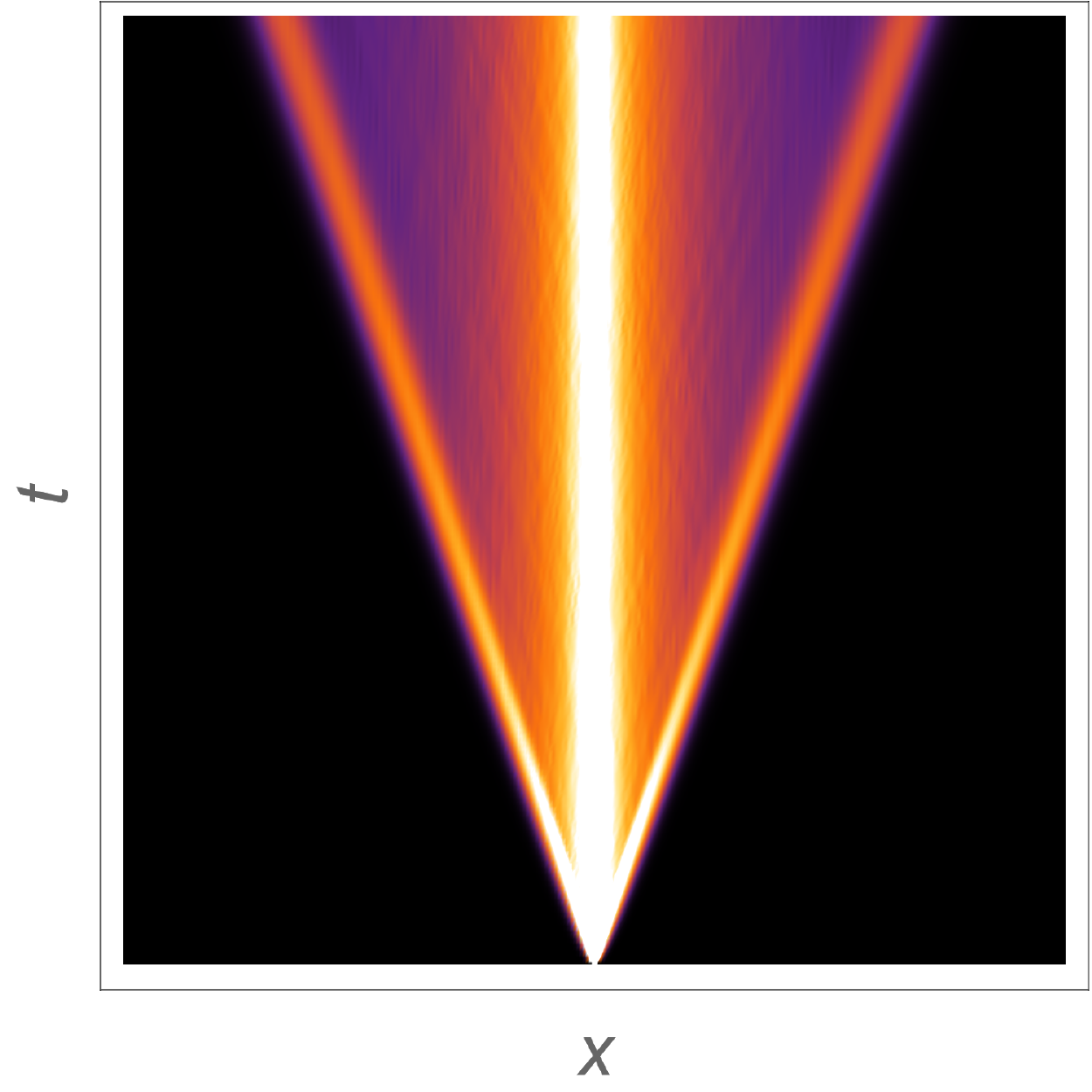}
\end{minipage}
\caption{\emph{Critical length-scales and dynamics in the strongly hyperuniform Ising model}. Left: the length-energy scaling of two relevant scales at the Ising critical point: the clean correlation length $\xi_{clean} \sim 1/|E|$ and the localization length $\xi_{loc} \sim 1/|E|^\alpha$. When $\alpha > 1$ the localization length diverges faster, so appears asymptotically larger than the correlation length in the low energy critical properties. The critical properties are then controlled by the clean Ising correlation length. Right: expansion of an initially localized wavepacket along the CII line at $\alpha = 2$. The bulk of the wavepacket is localized (corresponding to high energy components of the wavepacket), but there is a rapidly attenuating component that propagates at the light cone, depositing weight as it goes.} 
\label{fig1}
\end{center}
\end{figure}

\begin{figure}[b]
\begin{center}
\includegraphics[width = 0.45\textwidth]{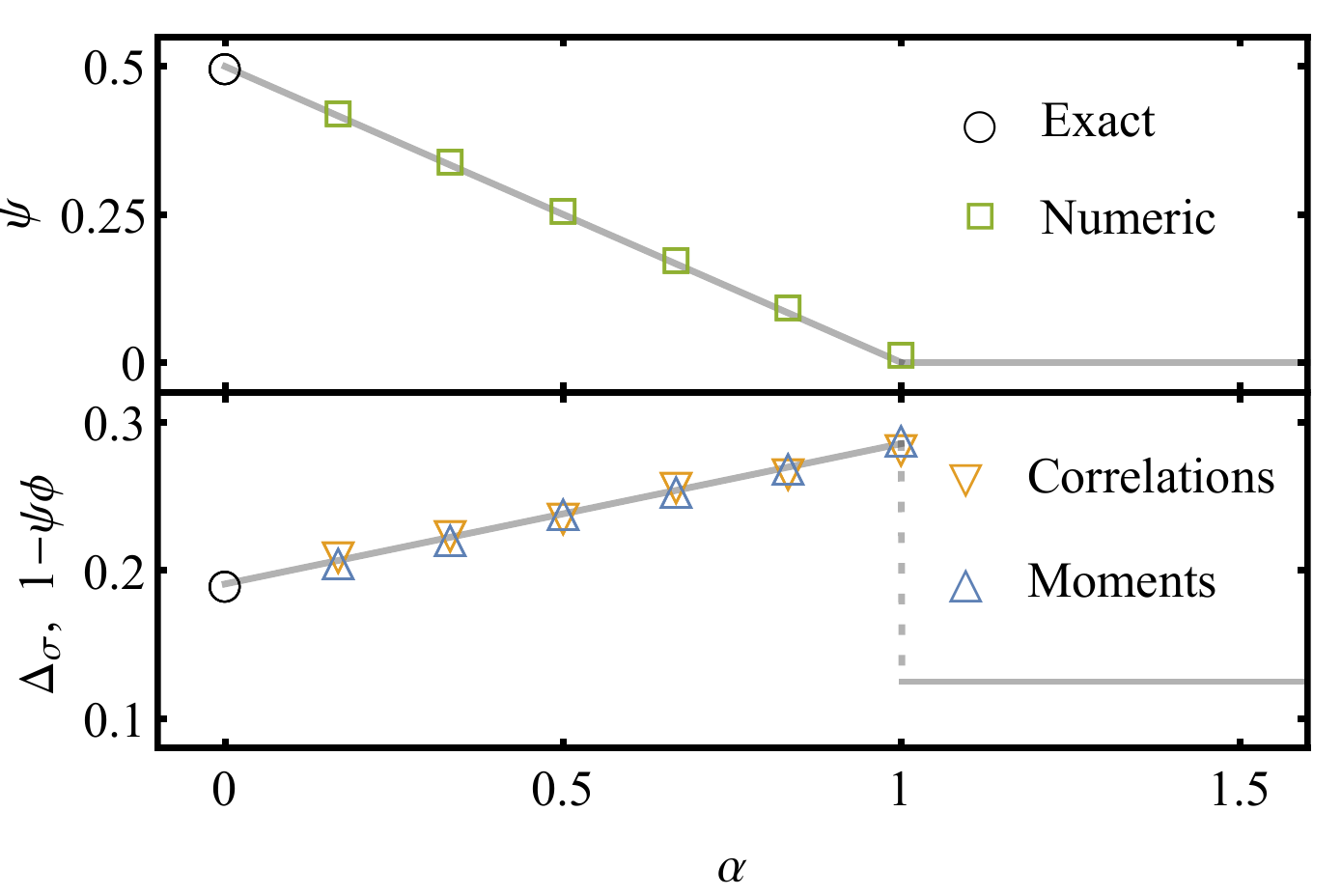}
\caption{\emph{Weakly hyperuniform systems}. Critical exponents at the Ising transition, vs. hyperuniformity parameter $\alpha$, extracted from the strong-disorder renormalization group. 
Upper panel shows the length-time scaling $\log t \sim l^\psi$, with the analytically exact results for $\alpha = 0$ \cite{fisher1992random} in agreement with the numerical results for $0 < \alpha \le 1$. 
Lower panel plots the \emph{average} order parameter scaling dimension, $\mean{\cexp{\tau^x_i \tau^x_{i + l}} } \sim l^{-2\Delta_\sigma}$, extracted from the correlations, and the scaling of the magnetic moment, $\mu_l \sim l^{\psi \phi}$.
The relation $\Delta_\sigma = 1 - \psi \phi$ is seen to hold.}
\label{whc}
\end{center}
\end{figure}

We explore the critical point (and near-critical phases) of the random hyperuniform TFIM as a function of a parameter $\alpha$, defined in Sec.~\ref{model}, that tunes the degree of hyperuniformity (Fig.~\ref{fig0}); $\alpha$ is related to the wandering exponent $\beta$ via $\beta = \max\{0, (1-\alpha)/2\}$. 
Our results are as follows. For strongly hyperuniform systems, disorder is irrelevant at the clean critical point, and (to leading order) does not affect thermodynamics or equal-time correlation functions. It is counter-intuitive that the disorder localise excitations, but remain irrelevant to critical proporties, this is possible as the localization length diverges faster than the clean correlation length (Fig.~\ref{fig1}) at low energies (or equivalently, under an RG flow). However, despite irrelevance, the dynamics is completely altered even for weak disorder: thermal transport vanishes, and autocorrelation functions do not decay to zero; a wavepacket has a ballistically moving front that remains well-defined at all times (a remnant of the $z = 1$ clean critical dynamics), but the front attenuates as it moves, and at late times the weight at the front shrinks to zero (Fig.~\ref{fig1}). We call the resulting unconventional critical point the ``critical Ising insulator'' (CII); because disorder acts as a dangerous irrelevant variable here, we are able to develop an essentially complete analytic understanding of the unusual dynamics on this critical line.

In the weakly hyperuniform case, disorder is relevant, and we find a line of infinite-randomness critical points with continuously varying critical exponents (Fig.~\ref{whc}). Unlike the uncorrelated $\alpha = 0$ case, there are no Griffiths phases for any degree of hyperuniformity; we explain this with an elementary counting argument. We explore these critical points via strong-disorder renormalization group (SDRG) methods. Our SDRG results for the average spin correlations at the critical point yield unexpected non-monotonic behavior: these correlations go as $\mean{\cexp{\x_i\x_{i+l}}} \sim l^{-2\Delta_\sigma}$, where the exponent $\Delta_\sigma$ first \emph{increases} as the model is made more hyperuniform, then drops discontinuously. We attribute this effect to rare regions (which dominate response in the conventional random TFIM) getting progressively less dominant, and eventually becoming subleading to typical regions.

The rest of this work is organized as follows. In Sec.~\ref{model} we introduce a family of Ising models with random hyperuniform couplings. In Sec.~\ref{pert} we use perturbative stability arguments, as well as exact results for the zero-mode wavefunction, to identify the perturbative (strongly hyperuniform) and nonperturbative (weakly hyperuniform) regimes. (In the process, we also generalize the Harris criterion to the hyperuniform case.) We then explore the equilibrium and dynamical properties of the strongly hyperuniform critical point (Sec.~\ref{SH}) and the weakly hyperuniform critical point (Sec.~\ref{WH}), using a combination of perturbative and strong-randomness methods. In Sec.~\ref{ED} we present numerical results, from exact diagonalization, on the evolution of correlation functions as the degree of hyperuniformity is changed. Finally, in Sec.~\ref{disc} we summarize our results and address their implications for more general phase transitions in hyperuniform systems.

\section{Models and realizations}\label{model}

We consider the transverse field Ising model (TFIM) with spatially varying couplings:

\beq
H = \frac{1}{2} \sum_{i=1}^L \left( h_i \z_i + J_i \x_i \x_{i+1} \right),
\label{eq:Ham}
\eeq
where $\tau^\alpha$ are the Pauli matrices. We construct the coefficients $h_i, J_i$ as follows. For concreteness, consider the $J_i$; we choose $J_i$ to have the form $J_i \equiv J_0 \exp(-\s q_i)$, where $q_j \equiv \frac{1}{\sqrt{L}} \sum\nolimits_k q_k e^{-i k j}$ (with $k = 2 \pi n/L$, and $n=1 \ldots L$), and $q_k$ are random numbers with correlations given by the structure factor $S_\alpha(k,k')$:

\beq\label{family}
S_\alpha(k, k') \equiv \mean{ q_k q_{-k'} } \sim |k|^\alpha \delta_{k k'},
\eeq
where from here on $\mean{\cdot}$ denotes disorder averaging. In numerics we use $q_k = |\sin(k/2)|^{\alpha/2}\frac{1}{\sqrt{L}}\sum_{j}\xi_j e^{i k j}$ ($j=1\ldots L$) for independently identically distributed (iid) $\xi_j$ drawn from the uniform distribution of mean $\mean{\xi_j}=0$ and unit variance $\mean{\xi_j^2}=1$. For this choice of $q_k$ one finds $\mean{ q_k q_{-k'} } = |\sin(k/2)|^{\alpha} \delta_{kk'} \sim |k|^\alpha \delta_{k k'}$ as required.

When $\s$ is small, we can expand $J_i$ to linear order in $q_i$, so that both have the same fluctuation properties; for the nature of the critical point, however, it is the distribution of $\ln J_i$ that we would like to be hyperuniform (Sec.~\ref{sec:zeromode}). It is known~\cite{torquato2016} that when $0 < \alpha < 1$, the fluctuations scale as $\std{\sum_{i = 1}^l q_i } \sim l^{(1-\alpha)/2}$, where $\std{\cdot}$ denotes the standard deviation; for $\alpha > 1$, the system is strongly hyperuniform, since these fluctuations are independent of the size of the region. 
Models with general $\alpha$ involve long-range correlations of the disordered couplings, as a result of their non-analytic behavior as $k \rightarrow 0$. For the bulk of our analysis and numerical work we fix $h_i$ to have the same distribution as $J_i$ in order to retain the standard self-duality properties of the Ising model. However, we have checked that our results are unaffected if, instead, we choose either $h_i$ or $J_i$ to be constant, so long as at least one of the terms is random and hyperuniform. 

Since these patterns~\eqref{family} have simple properties in Fourier space, they can in principle be implemented in systems of ultracold atoms using spatial light modulators (which engineer potentials in $k$-space~\cite{pasienski2008high, gaunt2012robust}). 
Spatial light modulators would allow one to realize hyperuniform couplings in, e.g., Rydberg-atom arrays, in which the TFIM has been realized~\cite{browaeys, schauss2015crystallization}. Also, in realizations of the TFIM that use quantum gas microscopes~\cite{simon2011}, all parameters can be addressed and tuned locally. Beyond ultracold gases, random hyperuniform couplings can also be straightforwardly realized in arrays of magnetic adatoms~\cite{nadj2014}, deposited precisely on surfaces using scanning-tunneling microscopy, which can be chosen to have random hyperuniform spacings.

For the specific case $\alpha = 2$, the structure factor is analytic at $k = 0$ so a simple local construction of the random potential for this case exists~\cite{gabrielli2004voronoi} (Fig.~\ref{fig0}, upper panel). Define $q_j = j + \xi_j$, where $\xi_j$ are iid random ``displacements'' with a mean $\mean{\xi_j}=0$. Then $J_j = J \exp(\s \{ q_j - q_{j-1}\})$. This choice of couplings is physically natural: it corresponds to exponentially decaying spin-spin interactions between spins on sites that are randomly displaced from equilibrium positions on a regular crystalline lattice.  
One can check that for weak variations in $J_i$, we have $S_\alpha(k,k') = \delta_{kk'} \mean{\xi^2}  \sin^2 \left(k/2\right) \sim \delta_{kk'} k^2$; thus, this model is indeed hyperuniform with $\alpha = 2$. Other hyperuniformity exponents $\alpha$ may be realised in a similar manner if the spin degrees of freedom are spaced such that the fluctuations on the number of spins in a given volume is hyperuniform. Such models may arise naturally for judiciously chosen processes~\cite{chertkov2016inverse,batten2008classical}.

We note that the TFIM with arbitrary couplings can be mapped via Jordan-Wigner transformation to a model of free Majorana fermions with spatially varying hopping. Specifically~\cite{ccl,Chandran:2017ab,crowley2018critical}, 

\beq\label{hm}
H = \frac{i}{2} \sum_i \left( J_i \gamma_{2i} \gamma_{2i+1} + h_i \gamma_{2i+1} \gamma_{2i+2} \right),
\eeq
where the Majorana operators are related to the spins via the relations

\beq
\gamma_{2i} \equiv \left( \prod\nolimits_{j < i} \tau^z_j \right) \tau^x_i; \quad \gamma_{2i+1} \equiv \left( \prod\nolimits_{j < i} \tau^z_j \right) \tau^y_i.
\eeq
This free-fermion representation allows for $H$ to be brought to a diagonal form $H = i \sum_n E_n \eta_{2n} \eta_{2n+1} $ using exact diagonalization, and thus permits studies of relatively large systems. 

\section{Critical points and phases vs. $\alpha$}\label{pert}

In this section we identify the various regimes of behavior as a function of the hyperuniformity parameter $\alpha$, using perturbative arguments and the exact solution for the zero mode of the Ising model. This leads us to separate the phase diagram into a regime where disorder is perturbatively irrelevant (i.e., for strongly hyperuniform couplings (Sec.~\ref{SH})) and a regime where it is relevant (i.e., for weakly hyperuniform couplings (Sec.~\ref{WH})). In subsequent sections we address these regimes separately, using the methods appropriate to each.

\subsection{Harris criterion} 

As a first step to understanding the relevance of hyperuniformity, we generalize the Harris criterion to random hyperuniform potentials in one dimension. The argument below generalizes that given by Luck~\cite{Luck:1993ad} for quasiperiodic potentials.

The control parameter $\delta = \mean{\log h_i - \log J_i}$ describes the deviation of a thermodynamic system from criticality. Analogously one can define a local control parameter $\delta_l$, which describes the deviation from criticality within a finite region of size $l$. The value of $\delta_l$ depends on the disorder realisation (or equivalently the choice of finite region); $\delta_l$ has mean value $\mean{\delta_l}=\delta$, and fluctuations given by the corresponding standard deviation $\std{\delta_l}$. 

In the strongly hyperuniform case, the fluctuations of $\delta_\xi$ within a region of the size of the correlation length $\xi$, are of scale $\std{\delta_\xi} \sim 1/\xi \sim \delta^{\nu}$. For the clean criticality to be stable, we require that $\mean{\delta_\xi} > \std{\delta_\xi}$, i.e. $\delta > \delta^\nu$, as the critical point is approached ($\delta \to 0$). In this case, stability of the clean universality to hyperuniform disorder requires $\nu \geq 1$.  
Thus the stability of the clean TFIM critical point in one dimension (where $\nu = 1$) is marginal. For weakly hyperuniform systems, the fluctuations are of order $\xi^{-(1+\alpha)/2}$, so the Harris criterion accordingly gives $\nu \geq 2/(1+\alpha)$. Thus the clean TFIM is perturbatively unstable to weakly hyperuniform potentials, while strongly hyperuniform potentials are marginal. To see that strongly hyperuniform potentials are in fact \emph{irrelevant}, we turn to the exact solution for the zero mode of the TFIM, which can be computed for arbitrary potentials.

\subsection{Zero mode}\label{sec:zeromode}

\begin{figure}[tb]
\begin{center}
\includegraphics[width=0.75\linewidth]{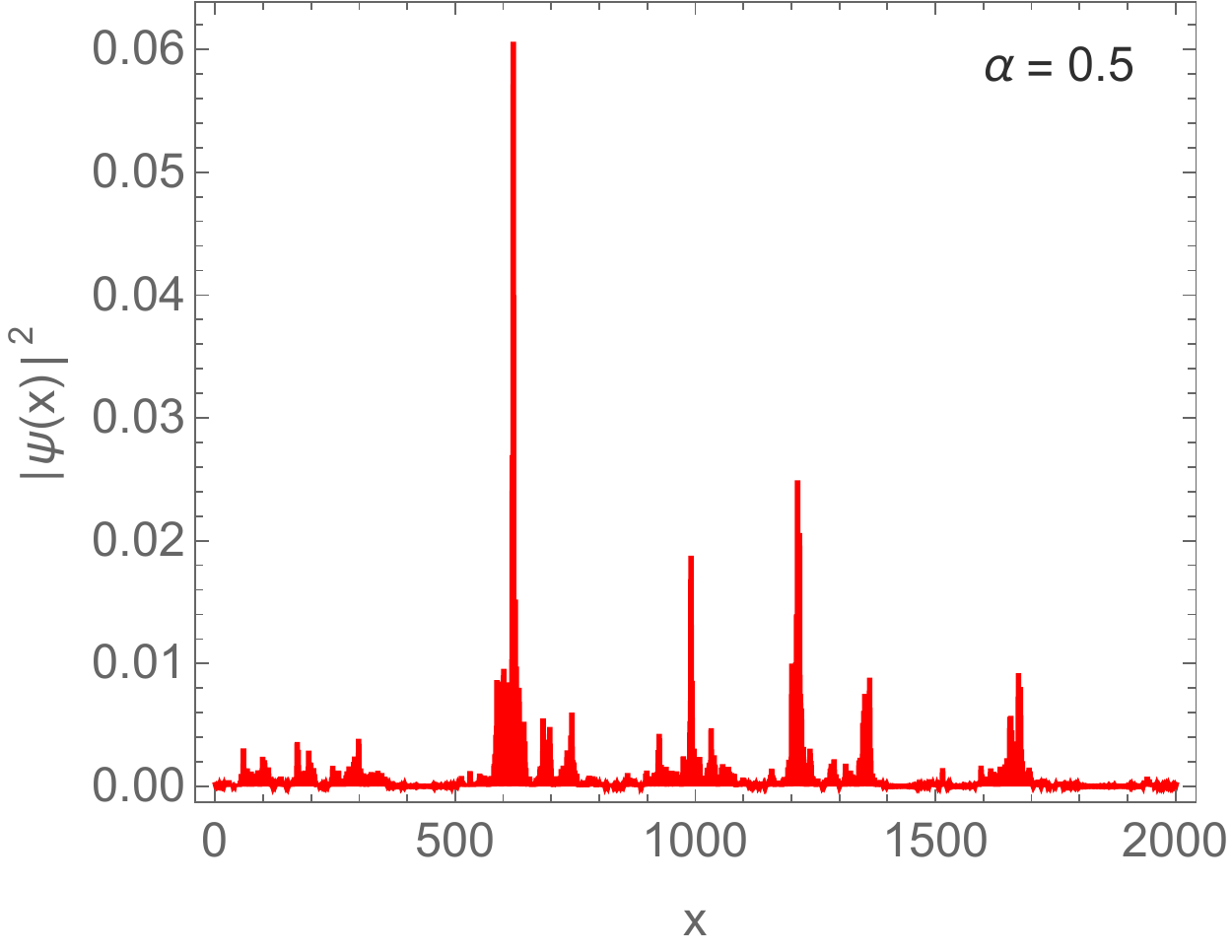}
\includegraphics[width=0.75\linewidth]{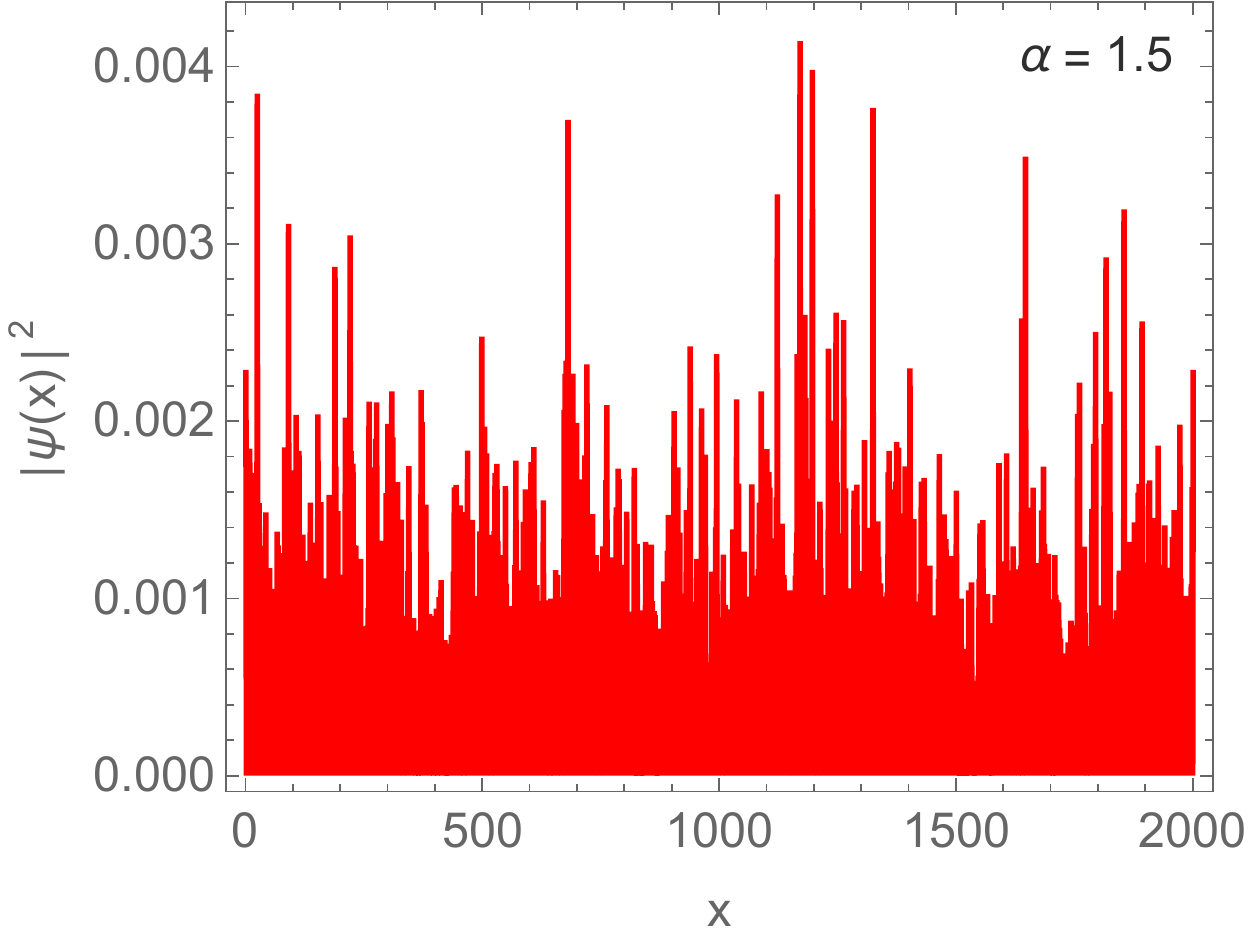}
\caption{Zero-mode wavefunction profiles $|c_i|$ (see Eq.~\eqref{zeromode}) in a representative sample in the weakly (left) and strongly (right) hyperuniform regimes.}
\label{zeromodefig}
\end{center}
\end{figure}

As a complementary way of probing the nature of the hyperuniform critical points, we use the following explicit construction of the zero mode of the critical Ising Hamiltonian~\cite{Chandran:2017ab}:
\beq\label{zeromode}
\eta_0 = \sum_i c_i \gamma_i, \quad c_i = \frac{1}{\mathcal{N}} \prod_{j < i} (h_j / J_j)
\eeq
where $\mathcal{N}$ is a normalisation factor.
Since strongly hyperuniform potentials do not cause this product to wander, the zero mode has uncorrelated random site-to-site fluctuations but no large-scale heterogeneity. 
For instance, in the $\alpha = 2$ model, $J_i = J \exp( \s \{\xi_i - \xi_{i-1}+1\})$, $h_i = h \exp(\s)$ so $c_i \propto (h/J)^i \exp(-\xi_i - \xi_1) \sim \exp(-\xi_i)$ at criticality. 
In the weakly hyperuniform case, by contrast, $\eta_0$ has strong amplitude fluctuations (Fig.~\ref{zeromodefig}), with sharp isolated peaks $c_j$. Moving a distance $l$ away from a peak, at criticality, the wavefunction amplitude typically decays as $c_{j+r} \sim \exp(-\text{const.} |r|^{(1-\alpha)/2})$. 
In the marginal case $\alpha = 1$, the product decays as $c_{j+r} \sim \exp(- \text{const.}\sqrt{\ln r})$ away from the peak $c_j$ i.e. slower than any power law. Therefore we expect the zero mode in this case to be spread out uniformly over the lattice, as in the strongly hyperuniform regime. 

\subsection{Energy-dependent localization length}

In the models we are considering here, states far from $E = 0$ are localized, \red{with a localization length $\xi$ given (at weak disorder) by the theory of weak localization
\begin{equation}
    \frac{1}{\xi(E_k)} \sim \rho(E_k)\left([J_k J_{-k}]+[h_k h_{-k}]\right) \sim \rho(E_k)S_{\alpha}(k,k)
    \label{eq:weak_loc}
\end{equation}
where $E_k$ is the clean dispersion, and $J_k,h_k$ are the appropriate Fourier components of the hyperuniform potential. To obtain this result consider a plane wave of momentum $k$, and treat the disorder perturbatively. The mean free distance is calculated by taking the product of the Fermi's golden rule decay time of the plane wave, and the group velocity. As the mean free distance is the only length scale in the problem we identify it with the localisation length, yielding Eq.~\eqref{eq:weak_loc}.}

A consequence of Eq.~\eqref{eq:weak_loc} is that the behavior of the localization length as $|E| \rightarrow 0$ is sensitive to $\alpha$. Specifically, if we begin at the clean critical point, and consider the weak-localization formula for $\xi$ as $|E| \rightarrow 0$, we find that $\xi \sim 1/|E|^{\alpha}$. This perturbative result is internally consistent whenever $k \,\xi(E_k) \gg 1$, where $k \propto E$ at the critical point; this is true for weak disorder when $\alpha \geq 1$, but breaks down as $|E| \rightarrow 0$ when $\alpha < 1$. Physically, in the strongly hyperuniform regime, the localization length diverges sufficiently rapidly at low energies that the momentum of an eigenstate $\sim |E|$ becomes asymptotically sharp compared to its momentum width $|E|^{\alpha}$, although the wavefunction is localized on the longest scales. By contrast, in the weakly hyperuniform regime, as with uncorrelated disorder, the perturbation theory breaks down at sufficiently low energies, and the low-energy localization properties are governed by nonperturbative effects.

\subsection{Stability of the critical Ising insulator}

The strongly hyperuniform case shares some features with the putative semimetal-to-metal critical point in disordered Weyl and Dirac systems~\cite{PhysRevB.33.3263, nhs, pixley_weyl, syzranov2018}. 
For Weyl systems, it is the zero-energy DOS rather than the spectrum of the disorder that vanishes at low energies; however, both mechanisms cause disorder to be perturbatively irrelevant as $|E| \rightarrow 0$. Nonperturbative rare-region effects destabilize Weyl semimetals in the presence of disorder, and one might wonder if some similar nonperturbative effect might arise as $|E| \rightarrow 0$ at the strongly hyperuniform critical point, leading it to flow to strong randomness. 
If some such scenario held, we would expect the true critical point to be at infinite randomness regardless of $\alpha$, and the structure of the zero mode to evolve smoothly with $\alpha$. 
But as we saw above, the exact zero mode in fact shows an abrupt change of behavior at the critical value $\alpha = 1$, supporting our case that there really is a sharp change in the critical properties between weakly and strongly hyperuniform regimes.

\section{Strongly hyperuniform case: ``critical Ising insulator''}\label{SH}

In this section we explore the critical behavior of the thermodynamics, equal-time correlations, transport, and dynamics when $\alpha > 1$. As noted already, the critical point has an energy-dependent localization length 

\beq\label{xi}
\xi(E) \sim 1/|E|^\alpha;
\eeq
this relation will be central to our analysis below. We first summarize the equilibrium properties of the critical point, which are (to leading order) unchanged by the hyperuniform potential; then turn to its dynamical properties, which are qualitatively different from those of the clean system. Finally, we extend our results away from the critical point.

\subsection{Equilibrium properties}\label{SH_eq}

\subsubsection{Density of states}

\begin{figure}
\begin{center}
\includegraphics[width=0.85\columnwidth]{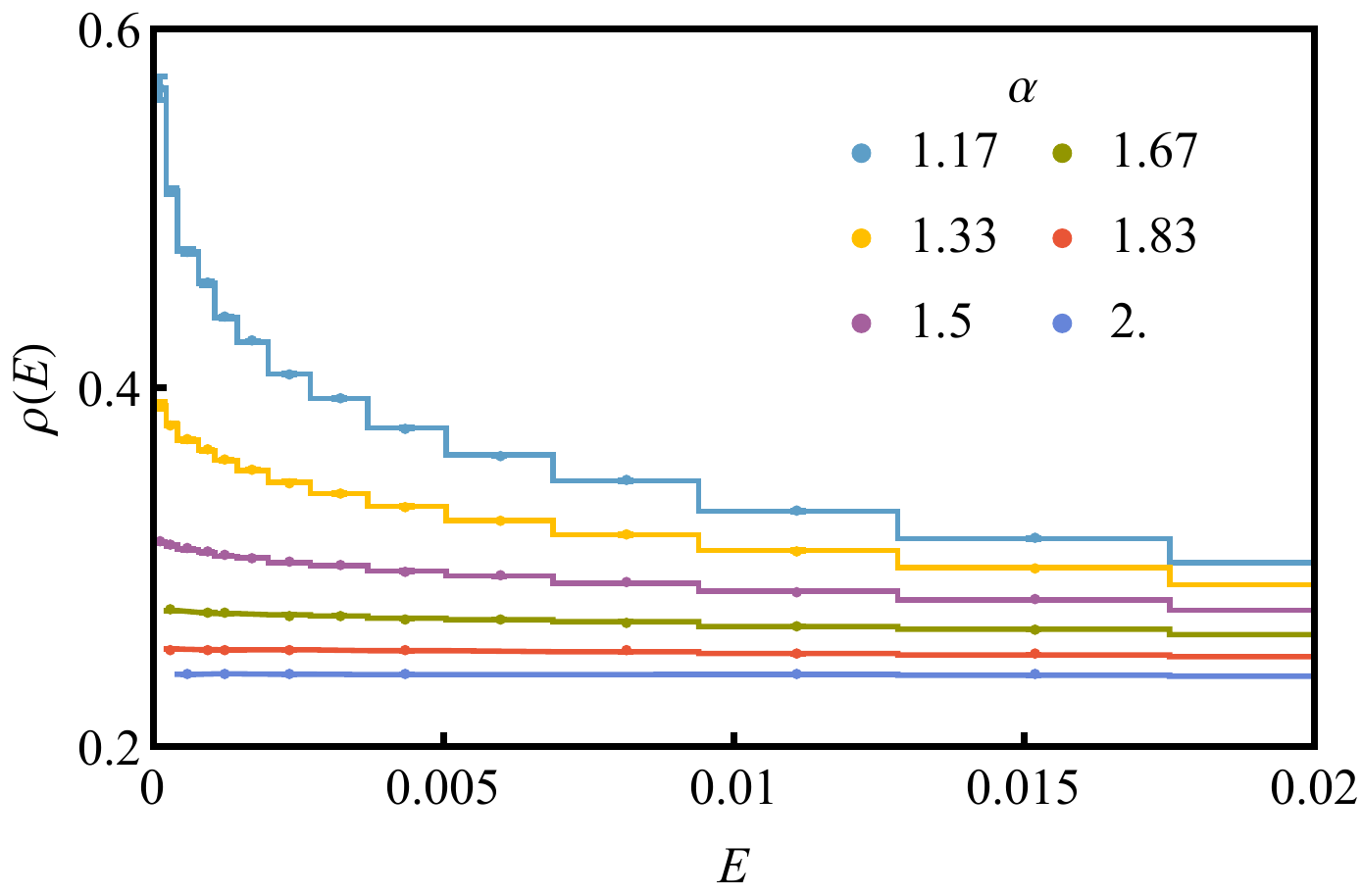}
\caption{
\emph{Density of states for strongly hyperuniform systems:} At low energies the density of states tends to a finite value as in the clean case. The hyperuniform modulation gives rise to a sub-leading correction in the form of a non-analytic cusp. The density of states was computed using the recursive method of Ref.~\cite{schmidt_dos} for parameters $L = 10^7$, $\s = 3/16$, and $n=40$ disorder realizations.
}
\label{Fig:StrongDOS}
\end{center}
\end{figure}


Since disorder is perturbatively irrelevant at this critical point, we expect the DOS of the disordered problem to approach a constant as $E \rightarrow 0$. However, there is a subleading non-analyticity in the DOS, for $\alpha \neq 2$. This non-analyticity follows from the non-analytic behavior of $\xi$; in fact, the two are related by the Thouless formula~\cite{thouless1972relation}:

\beq
\int dE' \big(\rho(E') - \rho_0(E')\big) \log|E - E'| \simeq \frac{1}{\xi(E)} \sim |E|^\alpha.
\label{eq:Thouless}
\eeq
where $\rho_0(E)$ is the DOS of the clean system. In general this equality requires $\rho(E) - \rho_0(E) \sim |E|^{\alpha - 1}$: thus, the nonanalytic dependence of $\xi^{-1}$ at low energies translates into a non-analyticity in the DOS. (The case $\alpha = 2$ is special: here, $\xi^{-1}(E)$ is an analytic function of $E$, so the nonanalytic DOS correction is absent there.) We see this nonanalytic behavior clearly by numerically evaluating the DOS for very large systems via the recursion method of Ref.~\cite{schmidt_dos} (Fig.~\ref{Fig:StrongDOS}). \red{Finally we note that due to the subleading nature of the correction to DOS we have neglected to precisely calculate logarithmic corrections in~\eqref{eq:Thouless}}.

\subsubsection{Equilibrium correlation functions}

On dimensional grounds, we expect equilibrium correlation functions at long distances to behave as they would for clean systems. The scale $\xi_l \sim 1/|E|^\alpha$ is much larger than $\ell_{\text{Ising}} \sim 1/E$. In the clean system, the correlations at a length-scale $l$ are set by wavefunctions at energies $E \sim 1/l$; however, these wavefunctions are only localized on much longer scales, so their localization properties are irrelevant for the equilibrium correlations. This expectation is consistent with the results of numerical simulations (Sec.~\ref{ED}).

\subsection{Dynamics at the critical point}

\subsubsection{Thermal transport}

Unlike equilibrium properties, transport is strongly modified by localization. The simplest conserved quantity in the Ising model is energy; accordingly, we focus on thermal transport. In the clean Ising chain, thermal transport is ballistic and the conductance is given by the appropriate Landauer formula, $\kappa \sim T$~\cite{rego_landauer}. This result no longer applies in the CII, but understanding how precisely energy is transported requires some care with the order of limits. In what follows, we consider a setup in which the Ising chain is connected to two leads at temperatures $T_{\rm L} \equiv T - \Delta T/2$ and $T_{\rm R} \equiv T + \Delta T/2$ respectively; we also make the linear-response assumption $\Delta T \ll T$.

Since $\xi \sim 1/|E|^\alpha$, in a chain of length $L$, excitations with $E \alt 1/L^{1/\alpha}$ are delocalized (and indeed ballistic). Since the level spacing scales as $1/L$, the \emph{number} of delocalized modes grows with system size, although the delocalized fraction of the spectrum decreases as $L^{-1/\alpha}$. There is a mesoscopic parameter regime for the temperature gradient such that $1/L \ll \Delta T \ll T \ll 1/L^{1/\alpha}$. In this mesoscopic regime, heat transport takes place through the delocalized states around zero energy and the clean-system Landauer result~\cite{rego_landauer} continues to apply. 

However, for thermodynamically long chains, this is not the appropriate order of limits. Instead, one keeps $\Delta T$ finite as $L \rightarrow \infty$, so that $1/L^{\alpha} \ll \Delta T \ll T$. In this limit, a vanishing fraction of the modes around $E = 0$ contribute to transport; moreover, the contribution of each delocalized mode is suppressed because it is effectively at very high temperature. The Landauer formula for the energy flux is~\cite{rego_landauer}:

\beq\label{landauer}
\dot{Q} = \frac{1}{2\pi} \int_0^\infty d\omega \omega \Big( n_{ \mathrm R}(\omega) - n_{\mathrm L}(\omega) \Big) t(\omega),
\eeq
where $n_{\rm R}, n_{\rm L}$ are the quasiparticle occupation numbers in the two leads, and $t(\omega)$ is the transmission coefficient of states at frequency $\omega$. The transmission coefficient is given by $\exp(-L/\xi) \sim \exp(-\alpha L E^\alpha)$, which we approximate by cutting off the integral at the energy scale $1/L^{1/\alpha}$ (this amounts to neglecting the exponentially suppressed transmission through localized states). The delocalized states with energies $E \alt 1/L^{1/\alpha}$ have occupation numbers that are effectively at high temperature since $1/L^{1/\alpha} \ll T$. Thus $n_{\mathrm R}(\omega) \sim 1/2 - \omega/T_{\mathrm R}$, and likewise for $n_{\mathrm L}$. Plugging these results into~\eqref{landauer} we find

\beq
\dot{Q} \sim \frac{\Delta T}{T^2} \frac{1}{L^{3/\alpha}} \Rightarrow \kappa \sim \frac{1}{T^2 L^{3/\alpha}},
\eeq
so the critical state is a thermal insulator, with a conductance that decays algebraically with chain length. 

\subsubsection{Wavepacket dynamics and autocorrelations}\label{wpdyn}

\begin{figure}[tb]
\begin{center}
\includegraphics[width = 0.4\textwidth]{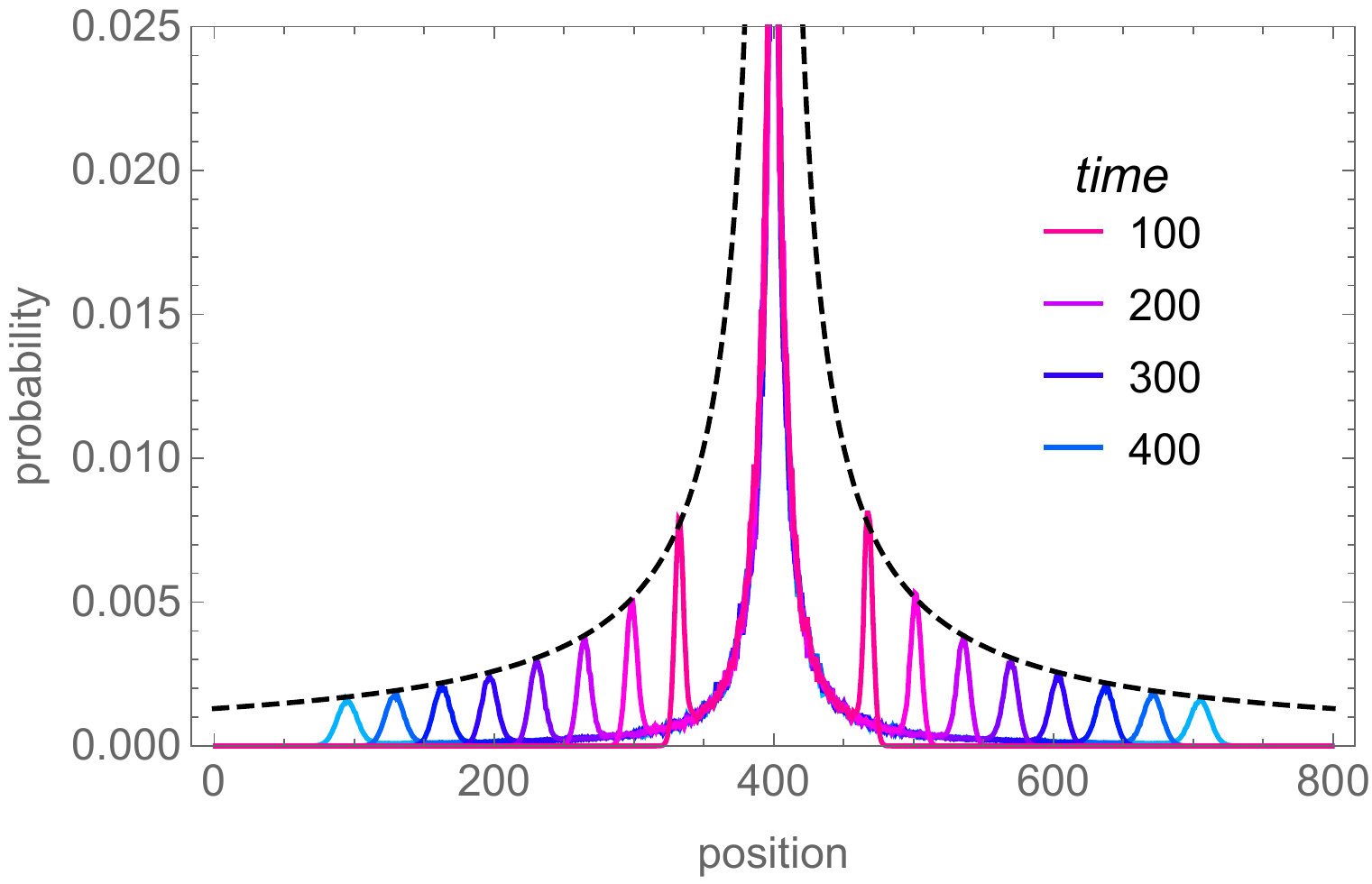}
\includegraphics[width = 0.4\textwidth]{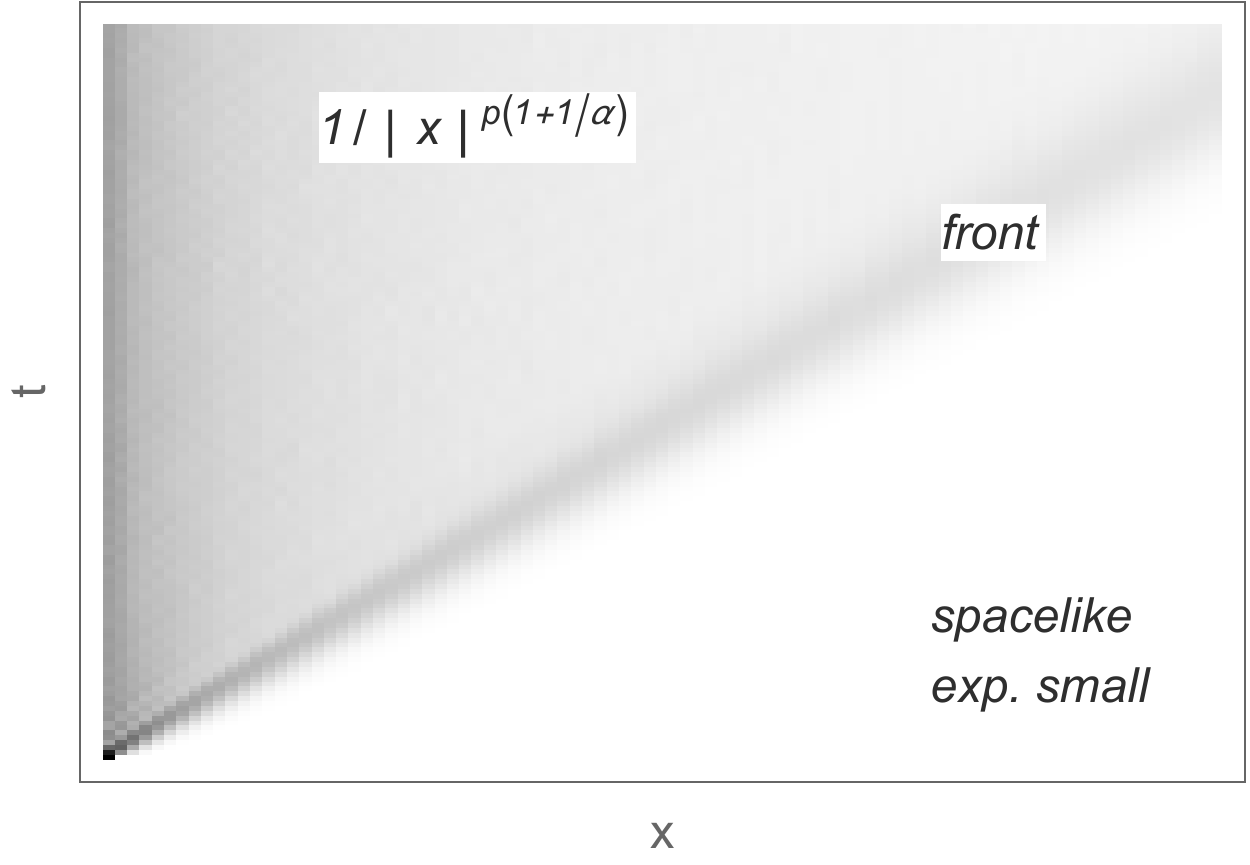}
\caption{Top: spreading of a wavepacket at the critical point for $\alpha = 2$. A well-defined ballistic front exists and moves out with the clean critical velocity; however, the weight at the front attenuates with time: its height shrinks as $1/t$ (dashed line) and it broadens as $\sqrt{t}$ (Supplemental Material App. A). Bottom: regimes of $\{ \gamma_{i+r}(t), \gamma_i(0) \}$. The autocorrelator is small outside the light-cone, grows to $r^{-2/\alpha}$ when $r \sim t$, then saturates at later times to a value $\{ \gamma_{i+r}(\infty), \gamma_i(0) \} = 1/r^{1 + 1/\alpha}$. Other correlation functions can exhibit multiples $p = 1, 2, \cdots$ of this basic power law.}
\label{frontspreading}
\end{center}
\end{figure}

We now turn to the behavior of autocorrelation functions and wavepacket dynamics; these quantities might be easier to probe, e.g., in ultracold atomic experiments, than transport. A particularly illuminating quantity to study is the dynamics of the Majorana fermion operator describing to the spreading of elementary excitations
\begin{equation}
    \gamma_i(t) = e^{i H t} \gamma_i e^{-i H t} = \sum_j U_{ij}(t)\gamma_j,
\end{equation}
the transition matrix is set by the anti-commutator $U_{ij}(t) = \{ \gamma_i(t), \gamma_j(0) \}$.
This quantity is closely related to the out-of-time-order correlator~\cite{LO, mss}. After addressing how elementary excitations spread, we turn to the behavior of general autocorrelation functions.

In the clean system at its critical point, operators spread ballistically. The situation in the disordered case is quite different. One can decompose the spatial Majorana degrees of freedom in terms of the fermionic eigenmodes $\gamma_i = \sum_n u_{in}\eta_n$.
After time evolution up to a timescale $t$, the projection of $\gamma_i$ onto modes that have localization lengths $\xi \alt t$ remains localized, while the rest of the operator moves ballistically to the light-cone, $r(t) = t$. Assuming the operator was initially spread out uniformly among modes, the fraction that is still spreading at time $t$ is $1/t^{1/\alpha}$. 
This spreading fraction consists of a well-defined but broadening peak, which is Gaussian in its outer tail, with height decaying as $t^{-2/\alpha}$ and width broadening as $t^{1/\alpha}$. 
As it moves, the front locally ``deposits'' intensity of order $t^{-1-1/\alpha}$ (one can see this from the conservation of total weight). When $\alpha > 1$ (i.e., in the CII), the height of the ``deposited'' operator is parametrically smaller than the height of the front, and the front remains well-defined at late times (Fig.~\ref{frontspreading}). 

Finally, the broadening of the front can be understood as follows. As we noted in Sec.~\ref{SH_eq}, the DOS in the presence of hyperuniform potentials gets modified to $\rho(E) \sim c + |E|^{\alpha - 1}$. Since the momentum of low-energy modes is asymptotically well-defined (because the localization length of a mode grows much faster than its wavelength), we can continue to associate a momentum to each eigenstate, and therefore interpret the DOS shift as providing an effective dispersion relation of the form $E(k) \sim a |k| + b |k|^\alpha$. This causes wavepackets to spread, with a width $\delta r(t) \sim t^{1/\alpha}$, when $1 < \alpha \leq 2$; this is the behavior we observe numerically~\footnote{For $\alpha > 2$, the leading correction to the dispersion is the (analytic) quadratic term, therefore in this regime the wavepacket broadens diffusively, and the height of the front decreases as $t^{-1/\alpha - 1/2}$.}. These various scaling relations are summarized in Table~\ref{t1}. For all $\alpha$ this broadening parametrically exceeds the $t^{1/3}$ broadening in the clean Ising chain~\cite{Platini2005, motrunich2018, xu2018a, 2018arXiv180305902K, PhysRevB.96.220302}.

\begin{table}[tb]\caption{Scaling properties of $\{ \gamma_{i+r}(t) \gamma_i(0) \} $ in the strongly hyperuniform regime ($1 < \alpha \leq 2$).}
\begin{tabular}{|c|c|}
\hline\label{t1}
Quantity & Behavior \\
\hline
Front height & $t^{-2/\alpha} \sim r^{-2/\alpha}$ \\
Front width & $t^{1/\alpha} \sim r^{1/\alpha}$ \\
Late-time saturation value & $r^{-1 - 1/\alpha}$\\
\hline
\end{tabular}
\end{table}

Because the TFIM is a model of free fermions, the results above can be used to infer the dynamics of any local perturbation that preserves the Ising symmetry (i.e., does not involve Jordan-Wigner strings). The various regimes of behavior of spatio-temporal autocorrelation functions such as, e.g., the retarded transverse field autocorrelation function $\langle [\gamma_i(t) \gamma_j(t), \gamma_0(0) \gamma_1(0)]\rangle \Theta(t)$, can also be deduced from the structure of the Heisenberg operator $\gamma_i(t)$. In the TFIM, local operators locally create or eliminate some number of quasiparticles, and each of these quasiparticles behaves as discussed above. Space-time correlation functions exhibit a well-defined but rapidly attenuating light-cone, and the behavior inside the light-cone clearly indicates localization: the memory of local perturbations persists indefinitely. 
The regimes of behavior of local autocorrelation functions are sketched in Fig.~\ref{frontspreading}. If one fixes a distance $r$ and measures a generic correlation function $C(r,t)$, it has three regimes: (i)~at times before the light-cone passes through, the correlation function is small, as causality demands; (ii)~at a time $r \sim t$, the correlation function grows to a value that is power-law small in $r$; (iii)~at times $r \gg t$, it saturates to a parametrically smaller value that is also power-law small in $r$. The precise values of these exponents depend on the operator.

\subsection{Away from criticality}

\subsubsection{Localization length}

Away from criticality, it appears that all states are localized at weak disorder. For simplicity we consider the paramagnetic phase (though our results extend to the ferromagnetic phase by Ising duality). Here, the clean system is gapped, with a dispersion relation $E \sim \sqrt{\Delta^2 + k^2}$. We are primarily concerned with the localization properties near $k = 0$, which corresponds to $E = \Delta$. The density of states is $\rho(E) \sim 1/\sqrt{E - \Delta}$, and the effective velocity of an excitation at $k(E) \sim \sqrt{(E - \Delta) \Delta}$ is $v(E) \sim \sqrt{E - \Delta}$. At leading order in perturbation theory, we find that the mean free \emph{time} diverges in the strongly hyperuniform regime as $\tau(E) \sim (E - \Delta)^{1/2(1 - \alpha)}$. However, because the velocity vanishes as $E \rightarrow \Delta$, the mean-free path goes as $\xi(E) \sim (E - \Delta)^{1 - \alpha/2}$. Thus $\xi$ vanishes (and perturbation theory ceases to be controlled) near the bottom of the band for $1 \leq \alpha < 2$. We check numerically that $\xi$ does indeed decrease near the bottom of the band in Fig.~\ref{LocLengthDelta}. \red{Precisely, the pairs $\psi^n, E_n$ were obtained by exact diagonalisation of a periodic chain, and indexed $n$ by their rank order in energy. A localisation length $\xi_n$ is then assigned to each one by a least squares fit to the relationship
\begin{equation}
    \log \sum_i \left| \psi_i^n \psi_{i+r}^n \right| = r /\xi_n + \mathrm{cons.}
\end{equation}
for $-L/2 < r < L/2$. The ensemble averaged values of $1/\xi = [1/\xi_n]$ and $E = [E_n]$ are then plotted in Fig.~\ref{LocLengthDelta}.}

For $\alpha = 2$, the localization length remains finite at the bottom of the band, and for $\alpha > 2$ it diverges within perturbation theory. Even for $\alpha > 2$, however, rare local configurations of the potential might smooth out the square-root divergence of the DOS and thus prevent $\xi$ from diverging; this issue is outside the scope of the present work.

\subsubsection{Equilibrium correlations}

The equilibrium correlation length at a distance $\delta$ from criticality is governed by the slowest-decaying modes in the system. The very lowest-energy modes are potentially tightly localized, in which case they cannot transmit correlations beyond their localization length; however, modes at an energy $\Delta E \sim \delta$ from the bottom of the band are still ``critical'' regime, so their localization lengths are parametrically larger than their inverse momenta, provided $\delta$ is sufficiently small. Thus the correlation length of the system is governed by these modes, which decay on a length-scale set by the distance from criticality, \emph{not} the localization length. Thus, up to potential numerical factors, the equilibrium behavior of the model away from criticality should be identical to that of the clean system. The minima in $1/\xi(E)$ which govern this behaviour are visible in Fig.~\ref{LocLengthDelta}.

\subsubsection{Transport and dynamics}

\begin{figure}[tb]
\begin{center}
\includegraphics[width = 0.4\textwidth]{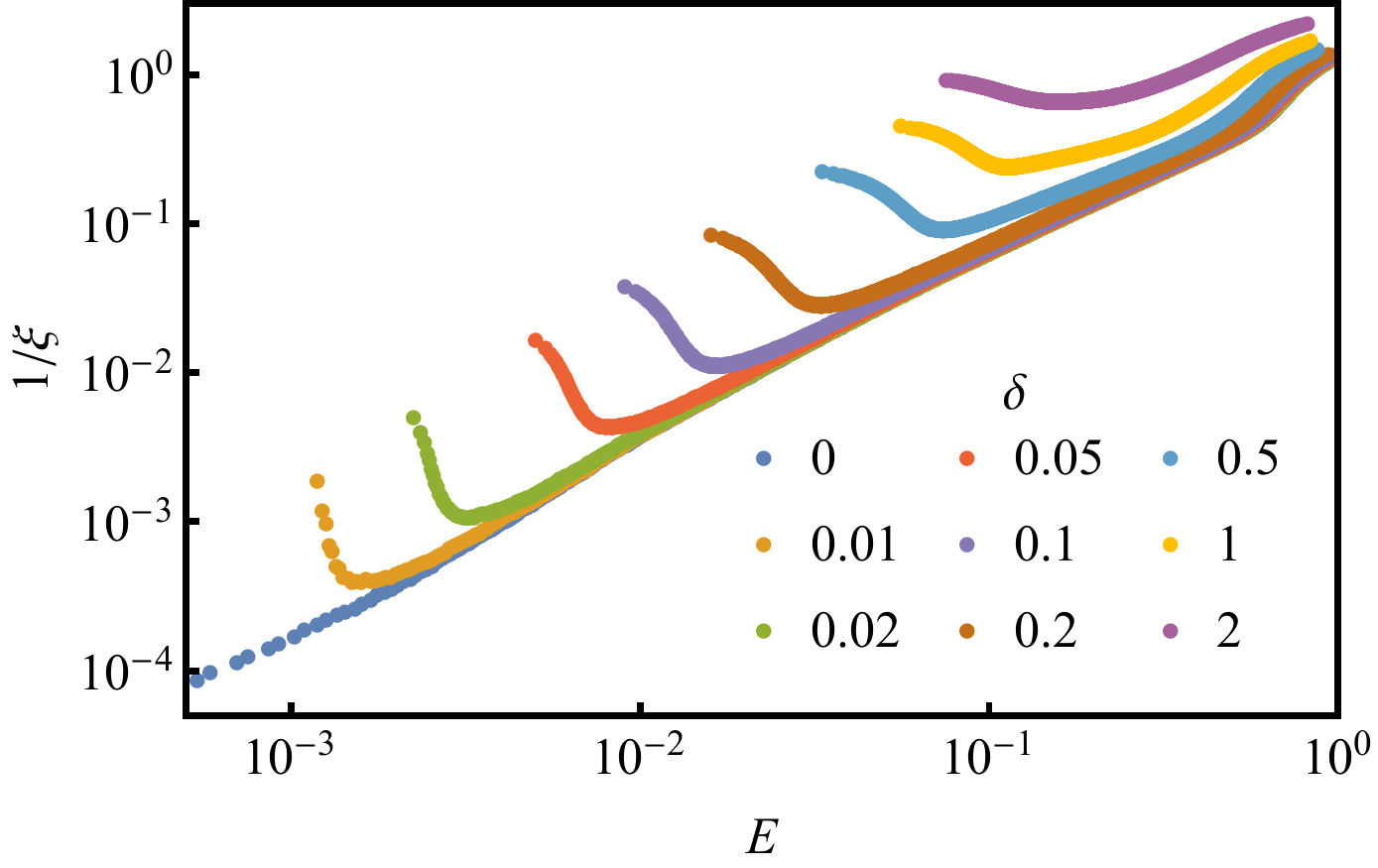}
\caption{
\emph{Minimum in the inverse localisation length $1/\xi(E)$:}
Disorder averaged $1/\xi$ is plotted versus $E$. At criticality the localisation length $\xi$ is monotonically increasing with decreasing excitation energy $E$. When one detunes away from criticality by $\delta$, the localisation length is minimised at an energy $E =O(\delta)$. Data for system sizes $L=5000$, $\alpha = 1.5$, $s=3/16$ error bars smaller than plot points.
}
\label{LocLengthDelta}
\end{center}
\end{figure}

Unlike equilibrium correlations, transport and dynamics are strongly affected by the localization properties of the model. For $\alpha \leq 2$ all modes are localized, with a non-diverging $\xi$. Wavepackets do not travel to infinity, and the finite-temperature thermal transport coefficients are \emph{exponentially} small in system size (in addition to being thermally activated). Specifically, at a distance $\delta$ from criticality, the least localized modes are those with $\Delta E \sim \delta$ above the gap, which have localization length $\xi \sim 1/\delta^\alpha$ (see Fig.~\ref{LocLengthDelta}). 
The conductance through a system of length $L$ is therefore suppressed as $\exp(-L \delta^\alpha)$.

\section{Weakly hyperuniform case: infinite-randomness critical line}\label{WH}

While perturbation theory about the clean limit allowed us to extract the behavior of physical observables in the strongly hyperuniform case (even when this behavior was drastically different from the clean system), such a perturbative approach evidently fails when disorder is relevant at the critical point. We approach this regime instead using strong-disorder renormalization-group (SDRG) methods and estimates based on counting rare regions. We first discuss the behavior of the density of states at the critical points, and the absence of Griffiths phases for $\alpha > 0$, as these can be understood using elementary counting arguments. We then present SDRG results for the evolution of critical exponents with $\alpha$. This section focuses on static properties, as these are the most directly accessible; however, we expect the dynamics throughout this phase to be qualitatively similar to that at the conventional infinite-randomness critical point. 

\subsection{Density of states}

\begin{figure}
\begin{center}
\includegraphics[width=0.85\columnwidth]{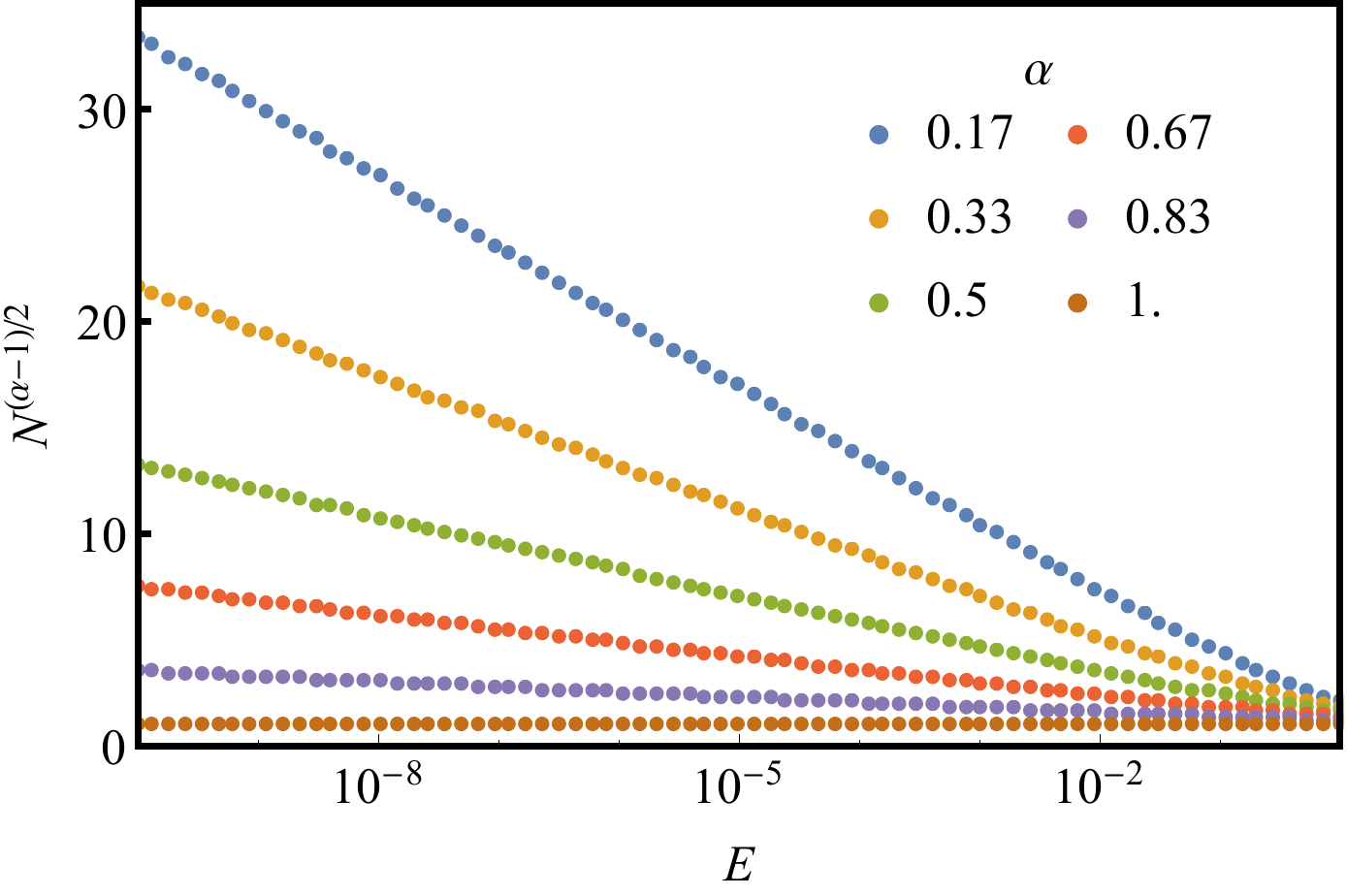}
\caption{
\emph{Density of states for weakly hyperuniform systems:} At low energies the integrated density of states scales as $N(E)^{(\alpha-1)/2} = c_1 \log E + c_0$ (for constants $c_0,c_1$), generalising the familiar $N(E) \sim 1/\log^{2} E$ of the iid ($\alpha=0$) case. Mean values 0f $N(E)$ are calculated by disorder averaging; statistical error is shown by error bars. Parameters: $L = 10^7$, $\s = 3/16$.
}
\label{Fig:WeakDOS}
\end{center}
\end{figure}

The thermodynamic properties of the critical point are captured by the density of states (DOS) near zero energy; this quantity can be estimated by adapting Ref.~\cite{eggarter}. The key result of that work is that, for a random hopping model, the integrated DOS up to energy $E$, $N(E) = \int_{0}^E \rho(E') dE'$, obeys the relation $N(E) \sim 1/2\ell(E)$, where $\ell$ is the spatial scale over which the quantity $\log[\prod\nolimits_j J_j/h_j]$ changes by an amount $\sim \log E$. For uncorrelated randomness, $\ell \sim \log^2 E$, leading to the familiar Dyson singularity in the DOS. For strongly hyperuniform potentials, when the potential is weak enough, the wandering does not grow with distance at all, so $\ell = \infty$, and the DOS is (to leading order) unaffected by weak randomness (but as discussed in Sec.~\ref{SH}, there are subleading non-analyticities). For $\alpha < 1$, $\ell \sim \log^{2/(1-\alpha)}(E)$ and therefore $\rho(E) \sim 1/(E \log^{1 + 2/(1-\alpha)} E)$. 
The corresponding low energy behaviour of the integrated DOS for weakly hyperuniform disorder, $N(E)^{(\alpha-1)/2} = c_1 \log E + c_0$ (for some constants $c_0,c_1$), is verified in Fig.~\ref{Fig:WeakDOS}. 
This construction of the DOS also gives an implicit relation between length- and timescales, $\ell(E)$. In the strongly hyperuniform case, $\ell(E) = \infty$, so randomness does not affect the dynamic critical exponent. In the weakly hyperuniform case, $\ell(E) \sim \log^{2/(1-\alpha)}(E)$, suggesting infinite-randomness behavior. 
For $\alpha = 1$, $\sqrt{\log \ell} \sim \log E$, so $\ell \sim \exp(\text{const.} \log^2 E)$. Thus, $\rho(E) \sim e^{-\text{const.} \log^2 E} \log E/E$, which vanishes at small $E$. This is subleading to the perturbative effects discussed in the previous section (as we would expect, since $\alpha = 1$ is in effect strongly hyperuniform).

\subsection{Griffiths effects}\label{griff}

The infinite-randomness critical point at $\alpha = 0$ is associated with ``Griffiths'' regimes on either side; in these regimes, the response to perturbations is dominated by rare regions that are in the wrong phase, and these contributions are parametrically dominant over the response from typical regions. Thus, for instance, in the paramagnetic phase sufficiently near the transition, the magnetization $m(h) \sim h^\gamma$ with $\gamma < 1$. This behavior occurs because the paramagnet contains an exponentially small (in size) density of regions that are locally in the ferromagnetic phase, and these regions have an exponentially large contribution to the susceptibility. These two exponentials combine to give a continuously varying power law, depending on the density of Griffiths regions, and the power law is less than one close to the critical point.

For $\alpha > 0$, there are no Griffiths regimes. One can see this by estimating the number of rare ferromagnetic regions in the microscopic model (i.e., regions of size $l$ for which the local control parameter $\delta_l$ exceeds some threshold $\beta$). To do this we need the probability distribution $P_L(\delta_l)$ for a region of size $l$. It is simpler to work with the characteristic functiuon

\beq
F_l(t) \equiv \mean{ \exp\left(\frac{i t}{l} \sum_{j =1}^l \ln\left\{ \frac{h_j}{J_j} \right\} \right) } \sim \exp\left(-t^2 l^{-(\alpha+1)}\right),
\eeq
where we used the Gaussian (though correlated) nature of the distribution of $\ln(h_i / J_i)$. Inverting the Fourier transform, we find

\beq
P_l(\delta_l) \sim \exp\left(-\delta_l^2 l^{1+\alpha}\right). 
\eeq
Thus the probability of a ferromagnetic region is suppressed faster than exponentially in $l$, whenever $\alpha > 0$. (In the strongly hyperuniform regime, it vanishes as a Gaussian in $l$, which is natural since in this regime the entire variance must come from the edge spins, which have a Gaussian distribution.) 

We now estimate the contribution to the susceptibility from these regions. At a field $h$, the ferromagnetic regions of size $\agt \log h$ fully magnetize. The density of such regions is given by $\exp(-(\log h)^{1+\alpha})$. This vanishes faster than a power law at small $h$, so it is always subleading to the paramagnetic response from typical regions. 

The arguments above applied only to the bare couplings; one might wonder if they continue to apply if one instead considers renormalized couplings. We argue below (Sec.~\ref{griffrg}), after introducing our renormalization scheme, that they do apply.

\subsubsection{Edge-spin susceptibility}

\begin{figure}[tb]
\begin{center}
\includegraphics[width=0.85\columnwidth]{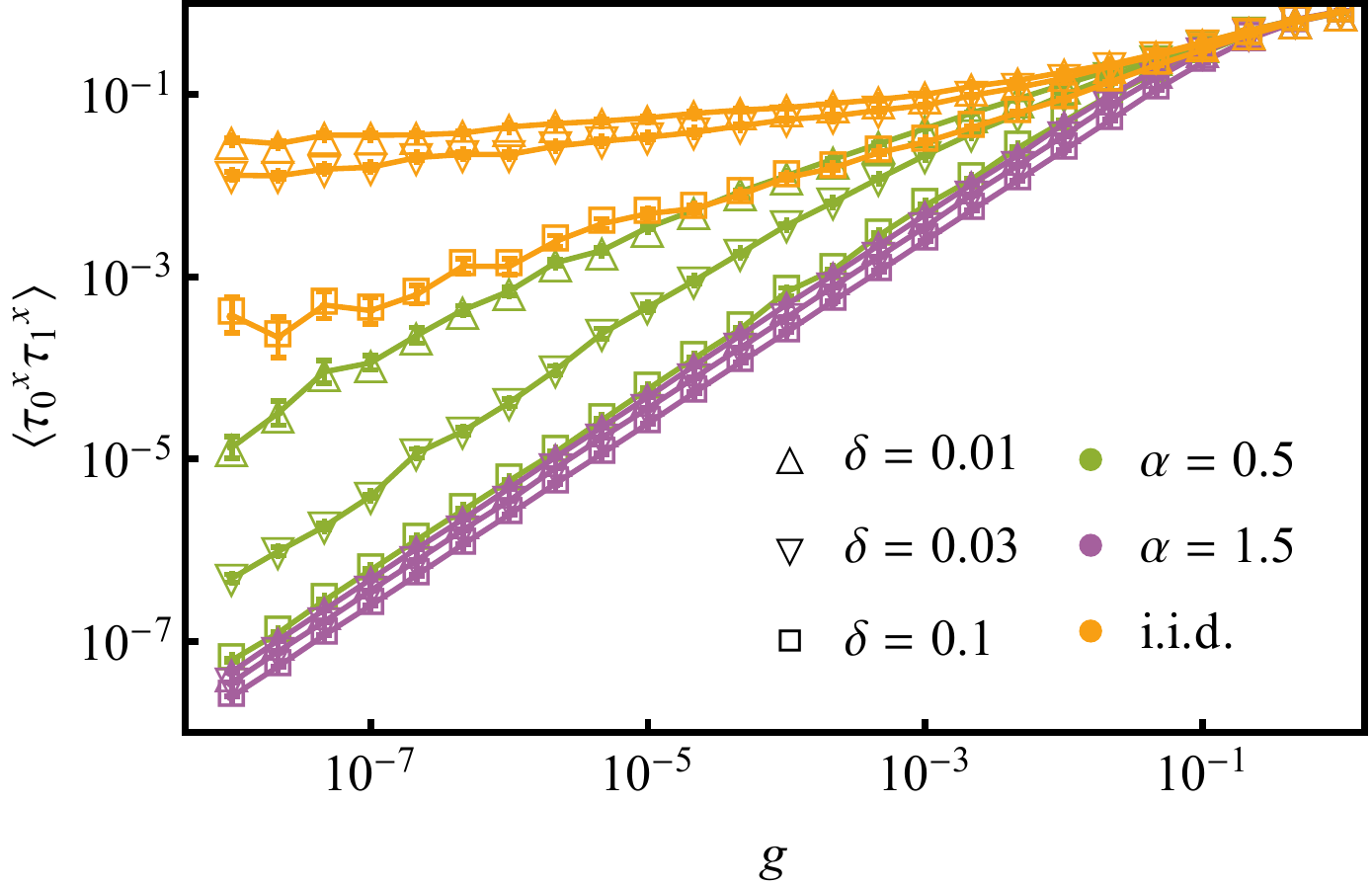}
\caption{
Susceptibility of a spin at the edge of the chain, vs. distance from the transition $\delta$, for the cases $\alpha = 0$ (independent random couplings), $\alpha = 0.5$ (weakly hyperuniform) and $\alpha = 1.5$ (strongly hyperuniform). Evidently even for weakly hyperuniform couplings, the susceptibility approaches linear behavior with field even very close to the transition, suggesting the absence of a Griffiths phase. Parameters: $L=2000$, $s=3/16$
}
\label{edgefig}
\end{center}
\end{figure}

This analytic argument against Griffiths effects is borne out numerically by studying the magnetization of the chain in response to a field applied at the edge; this is a simple way of computing a lower bound to the susceptibility of Ising chains (Fig.~\ref{edgefig}). To do this within the free-fermion description of the TFIM, one can introduce an artificial edge spin $\tau_0$ at one end of the chain, which couples to the leftmost spin via a coupling $g \tau^x_0 \tau^x_1$ but has no transverse field acting on it~\cite{mccoy_edge}. The susceptibility of the edge spin to this field is given by the $g$-dependence of the quantity $\langle \tau^x_0 \tau^x_1 \rangle$, which we can compute within the free-fermion theory. Our results are consistent with the absence of a Griffiths phase at $\alpha > 0$: the low-field susceptibility appears to be asymptotically linear in the field even very close to the critical point, in contrast to the $\alpha = 0$ case.

\subsection{SDRG and correlation functions}

\begin{figure}[tb]
\begin{center}
\includegraphics[width = 0.4\textwidth]{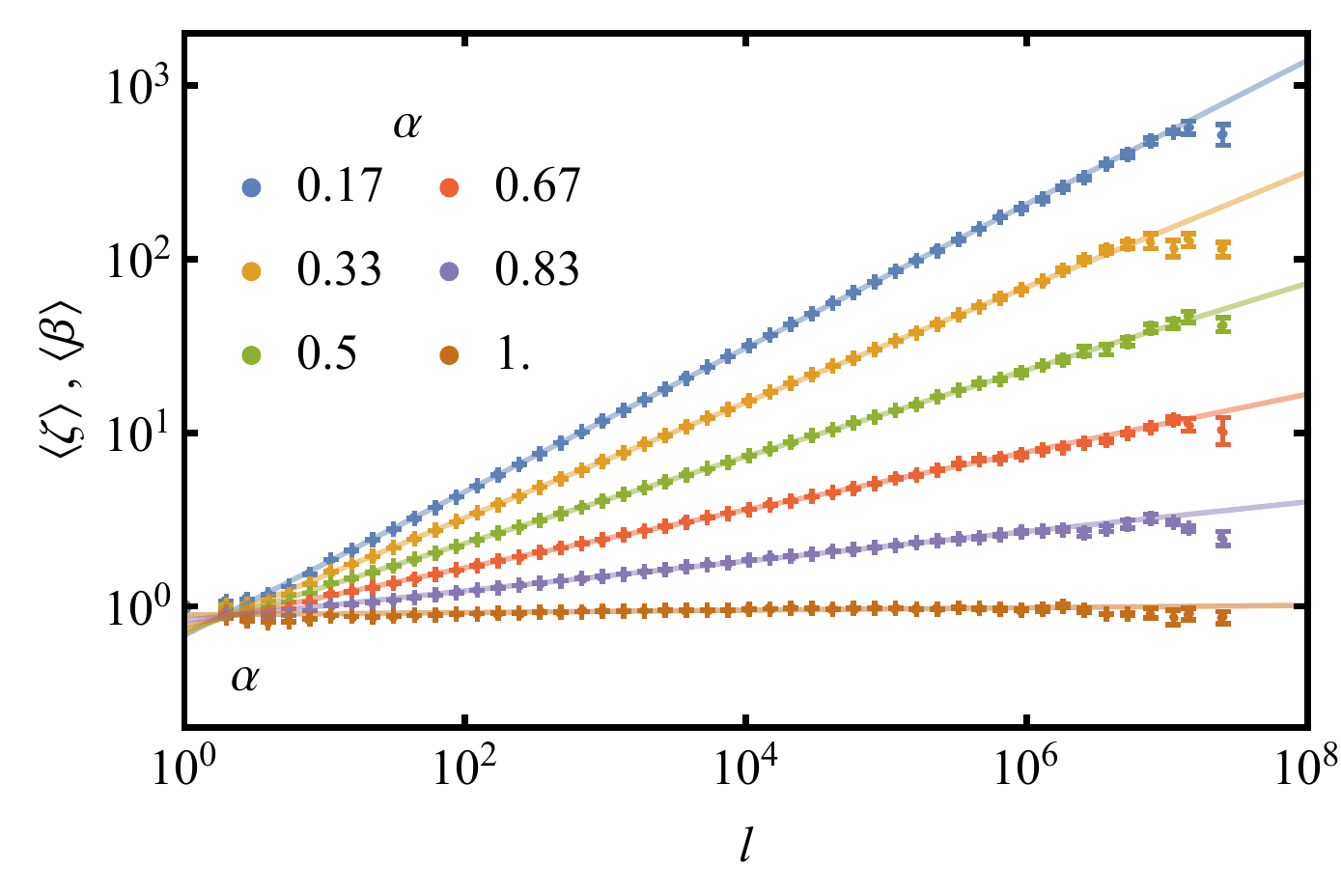}
\includegraphics[width = 0.4\textwidth]{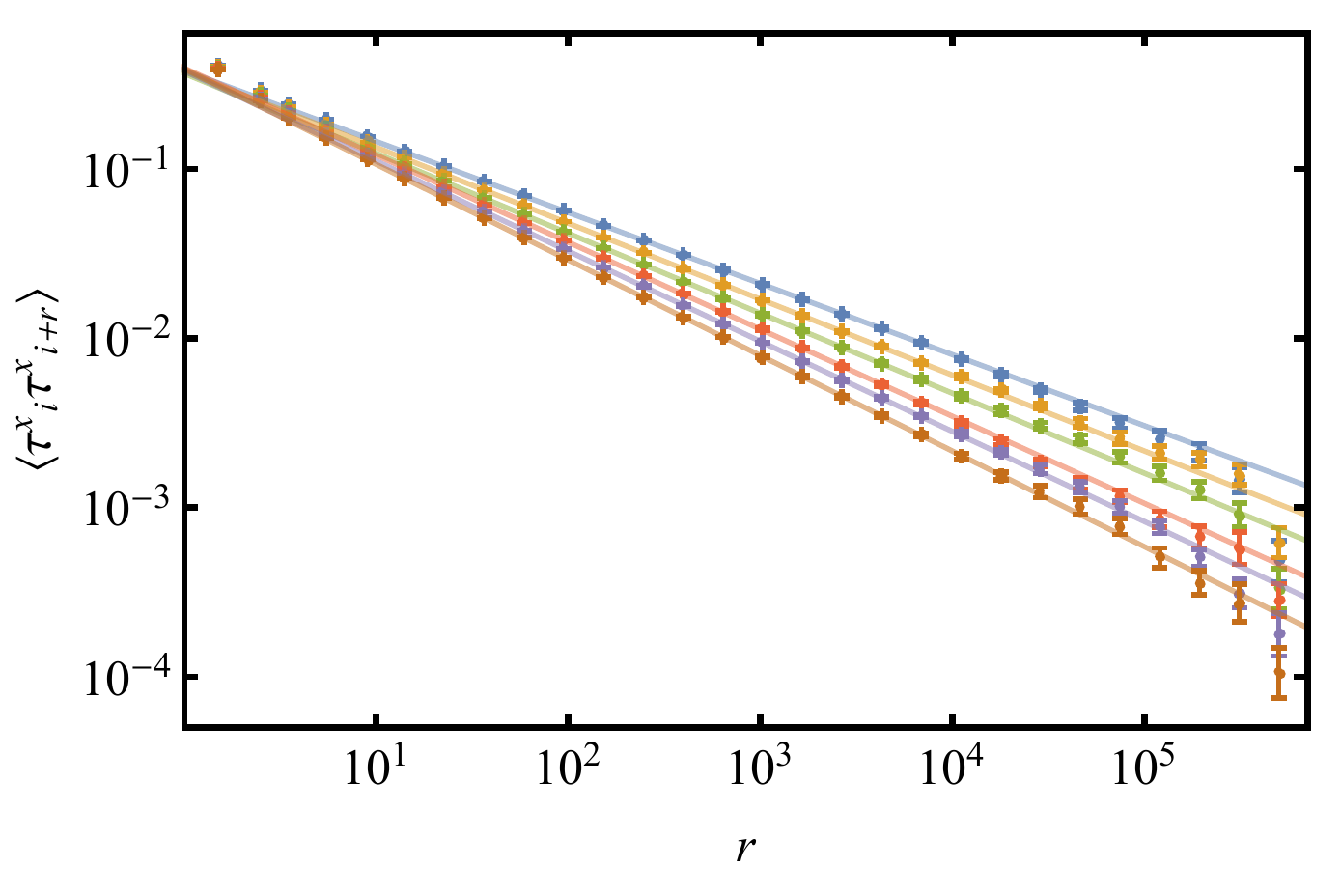}
\caption{
\emph{Strong disorder RG exponents:}
Upper panel: flow of typical coupling scale $\mean{\zeta},\mean{\beta}$ vs. length scale $l$ under SDRG, for various values of $\alpha$. The power law $\mean{\zeta},\mean{\beta}\sim l^\psi$ is exhibited with exponent $\psi = (1-\alpha)/2$. Lower panel: average correlator $C_{xx}(r) = \langle \tau^x_i \tau^x_{i+r} \rangle$ as a function of spacing. The decay is algebraic, and fits to the form $C_{xx}(r) \sim 1/r^{2(1-\psi\phi)}$ where the typical moment at scale $l$ scales as $\mu \sim l^{\psi\phi}$ (Scaling of moments is shown in the Supplemental Material App. C). Parameters: $L=10^8$, $s=3/16$}
\label{rsrgresults}
\end{center}
\end{figure}

To probe the nature of the critical point in the weakly hyperuniform case, we numerically apply the standard SDRG for the random TFIM~\cite{ma1979random, fisher1992random, fisher1995critical, fisher1999phase,  igloi2018transverse} to the hyperuniform case. The SDRG rules involve picking the largest coupling and eliminating it. If the largest coupling is a bond, one creates a new effective spin with transverse field $h_i h_{i+1}/J_i$; if the largest coupling is a field, one eliminates the corresponding spin to create a new effective bond $J_i J_{i+1}/h_i$. An important property of these rules is that the effective bonds at any stage in the SDRG are products of microscopic $J_i$ divided by products of microscopic $h_i$, and vice versa for the fields. If one runs the RG until the system size is rescaled by a factor $l$, the typical coupling scales as $\prod_{j=i}^{i+l} (J_j / h_j) \sim \exp(-\mathrm{const.}\,l^{(1-\alpha)/2})$. Thus the spacetime scaling at this critical point has the infinite-randomness form $t \sim \exp(\mathrm{const.}\, l^\psi)$, with $\psi = (1-\alpha)/2$. 

Other key exponents, such as the scaling of mean cluster moments, can be extracted from numerically iterating the SDRG rules for large systems (Fig.~\ref{rsrgresults}). The mean cluster moment $\mu$ at a length-scale $\ell$ goes as $\mu \sim l^{\psi \phi}$, where $\psi \phi$ decreases linearly from its $\alpha = 0$ value as $\alpha$ is increased. Thus, hyperuniformity yields sparser spin clusters than iid randomness. In the uncorrelated case, the mean cluster moment and the exponent that governs decay of mean order parameter correlations are related: $C_{xx}(r) \sim 1/|r|^{2\Delta_\sigma}$, where $\Delta_\sigma = 1 - \psi \phi$. We find numerically that this relation continues to hold for the weakly hyperuniform case (as one might expect, since the argument for this relation in Ref.~\cite{fisher1995critical} is quite general). 

A surprising implication of our results is that mean correlations at the critical point actually decay \emph{faster} in the weakly hyperuniform case than in the uncorrelated case (although the decay in the uncorrelated case is already faster than in the clean TFIM). Thus, as one tunes $\alpha$, it seems the exponent $\Delta_\sigma$ must first increase, and then discontinuously decrease to the clean Ising value at $\alpha = 1$. By contrast, the \emph{typical} correlations keep getting longer-ranged as the degree of hyperuniformity increases, going as $\exp(-|r|^{\psi})$. These observations can be qualitatively reconciled as follows: hyperuniformity involves local anticorrelations, which make spin clusters sparser than for independent randomness; therefore, correlations due to rare clusters are suppressed. At the same time, $\psi$ decreases so typical correlations become longer-ranged. Eventually, at $\alpha = 1$, typical regions begin to dominate over rare regions, and one enters the strongly hyperuniform regime. 

\subsubsection{Griffiths effects}\label{griffrg}

We now return to the question of whether Griffiths phases exist. This was already addressed for the bare theory~\ref{griff}; we now argue that renormalization does not change this basic conclusion. Consider running the RG out to some finite length-scale $\ell$; at this scale, the system consists of effective spins consisting of $O(\ell)$ microscopic spins, subject to transverse fields of the form $\tilde{h} = (h_k h_{k+1} h_{k+2} \ldots h_{k+\ell})/(J_{k+1} J_{k+2} \ldots J_{k+\ell})$, and coupled by bonds $\tilde{J}$ that are likewise products of adjacent $J$'s divided by products of adjacent $h$'s. Suppose we take a region of size $l \gg \ell$. The wandering in this region obeys the identity $\tilde{\delta} = \sum_{\alpha} \log(\tilde{J}_\alpha/\tilde{h}_\alpha) = \sum_i \log(J_i /h_i) = \delta$, where $\alpha$ labels the $o(l/\ell)$ effective spins and $i$ denotes the original microscopic spins in that region: this identity is an immediate consequence of the SDRG rules. Thus, if one terminates the RG after finitely many steps, the asymptotics of the wandering on much larger scales are unaffected, and the argument of Sec.~\ref{griff} goes through.

\section{Exact diagonalization results for correlation functions}\label{ED}

The previous sections addressed the properties of the weakly and strongly hyperuniform cases, using different methods (perturbation theory and SDRG respectively). In this section we discuss how correlation functions evolve as one tunes $\alpha$, using exact diagonalization. We first discuss two-point spin correlations, and then the ground state entanglement entropy.

\subsection{Spin-spin correlations}

The free fermion character of the Ising model allows us to perform simulations on systems of up to a few thousand sites. We focus on the equal-time correlation function of the order parameter, $C_{xx}(r) \equiv \langle \tau^x_i \tau^x_{i+r} \rangle$; this can be expressed as a determinant of free-fermion Green's functions~\cite{sachdev_book}. We have checked that correlation functions evolve qualitatively similarly. We set the disorder strength $\s = 3/16$: when the disorder is either much weaker or much stronger, we see strong transients. For weak disorder, these transients are expected, as the system is clean on short scales. At strong disorder, the states away from $E = 0$ are effectively site-localized and do not see the hyperuniformity; the universal regime of $\xi(E)$ shrinks to very small energies or equivalently to very large length scales.

\begin{figure}[tb]
\begin{center}
\includegraphics[width = 0.4\textwidth]{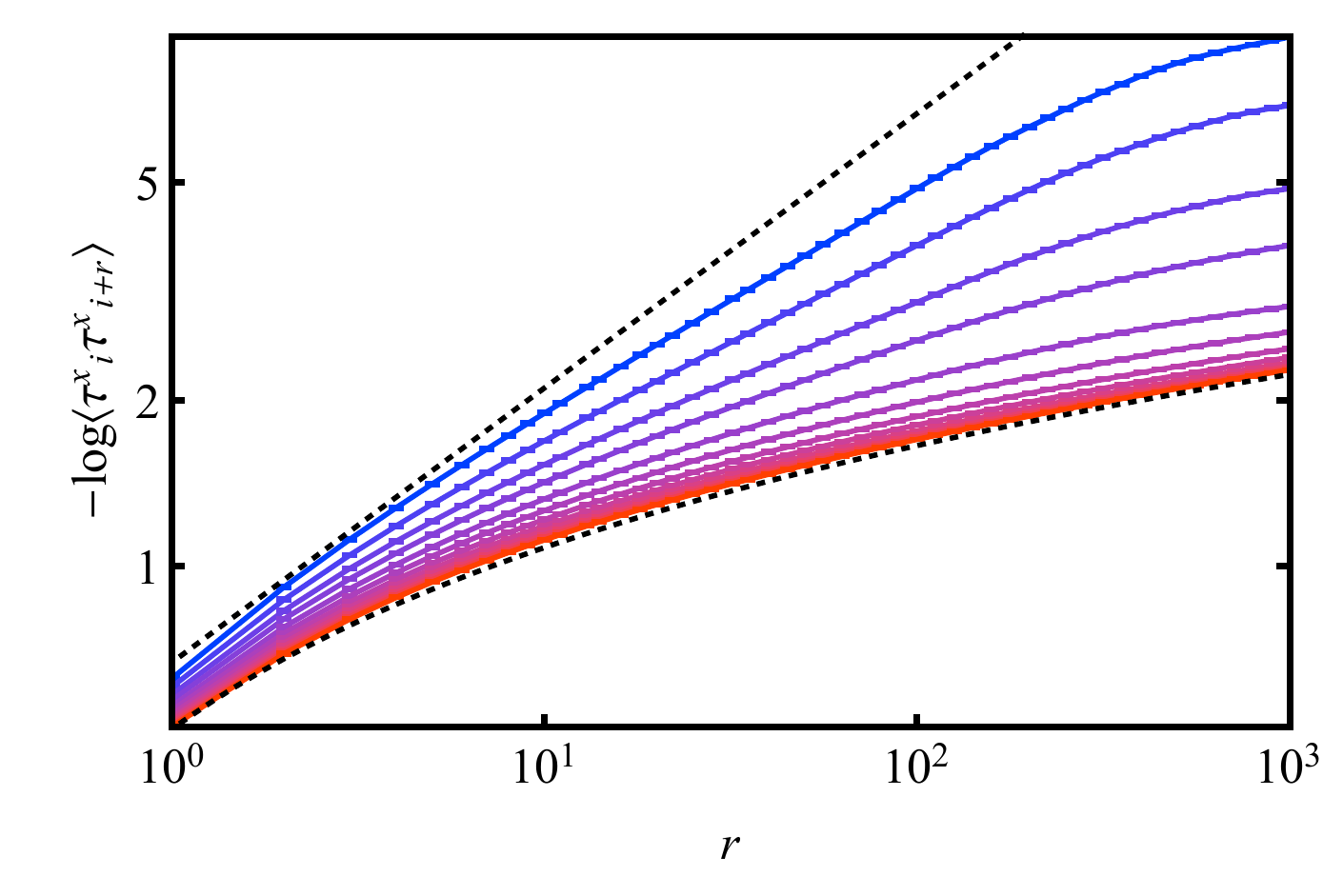}
\includegraphics[width = 0.4\textwidth]{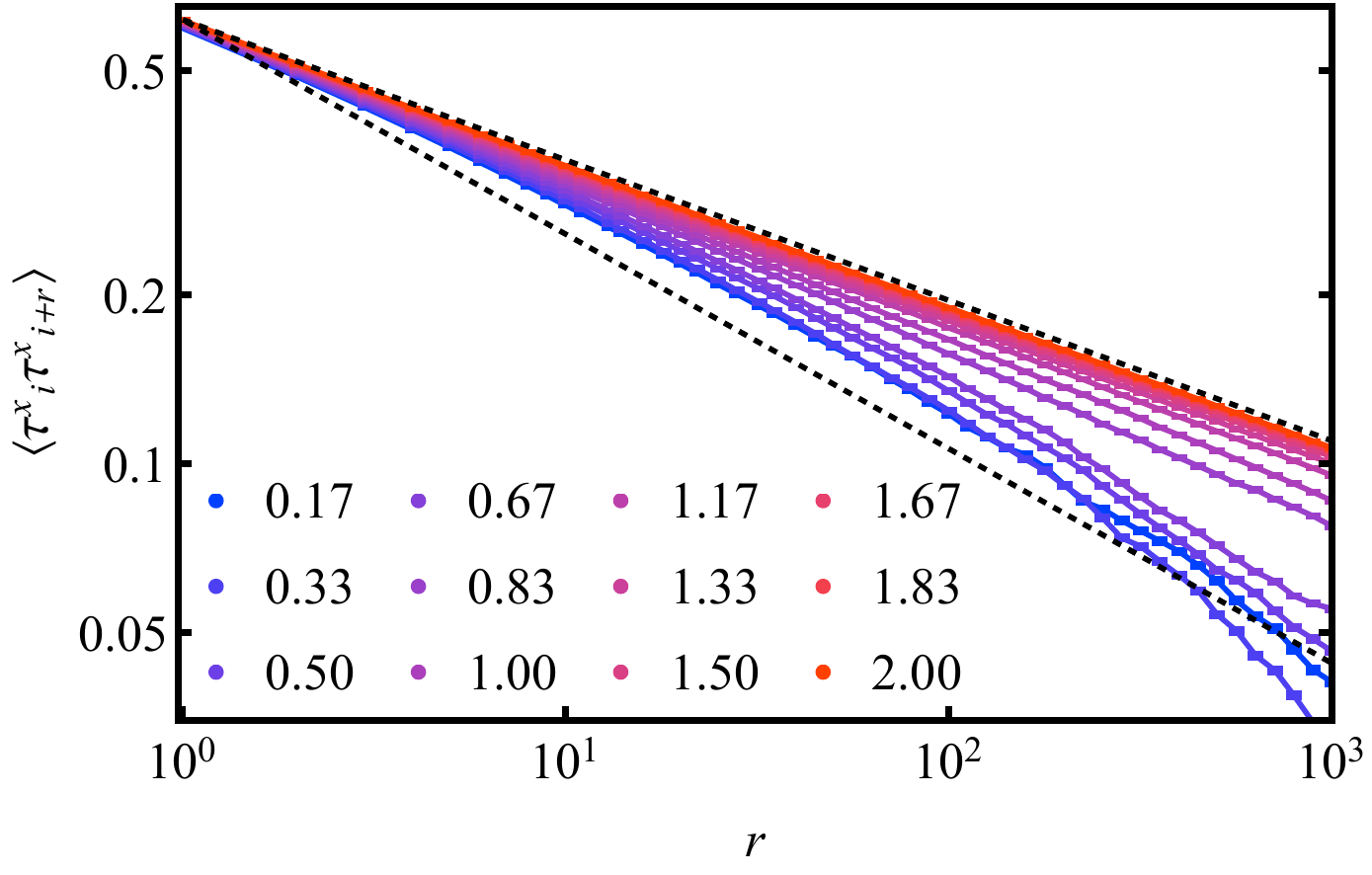}
\caption{Typical (upper) and mean (lower) order-parameter correlation functions $\cexp{\x_i\x_{i+r}}$ as a function of distance for different $\alpha$ (legend inset). The behavior of the typical correlation function is consistent with a power law in the strongly hyperuniform case ($\alpha > 1$) and with a stretched exponential in the weakly hyperuniform case ($0 < \alpha < 1$). The mean correlator decays with the clean Ising exponent in the strongly hyperuniform case, but clearly faster in the weakly hyperuniform case. In the weakly hyperuniform case we do not see a clean power law at large scales; it seems that our data here are still dominated by typical rather than mean behavior. Parameters: $L=2000$ periodic chain, $s=3/16$}
\label{edcorrs}
\end{center}
\end{figure}

\begin{figure}[t]
\begin{center}
\includegraphics[width = 0.4\textwidth]{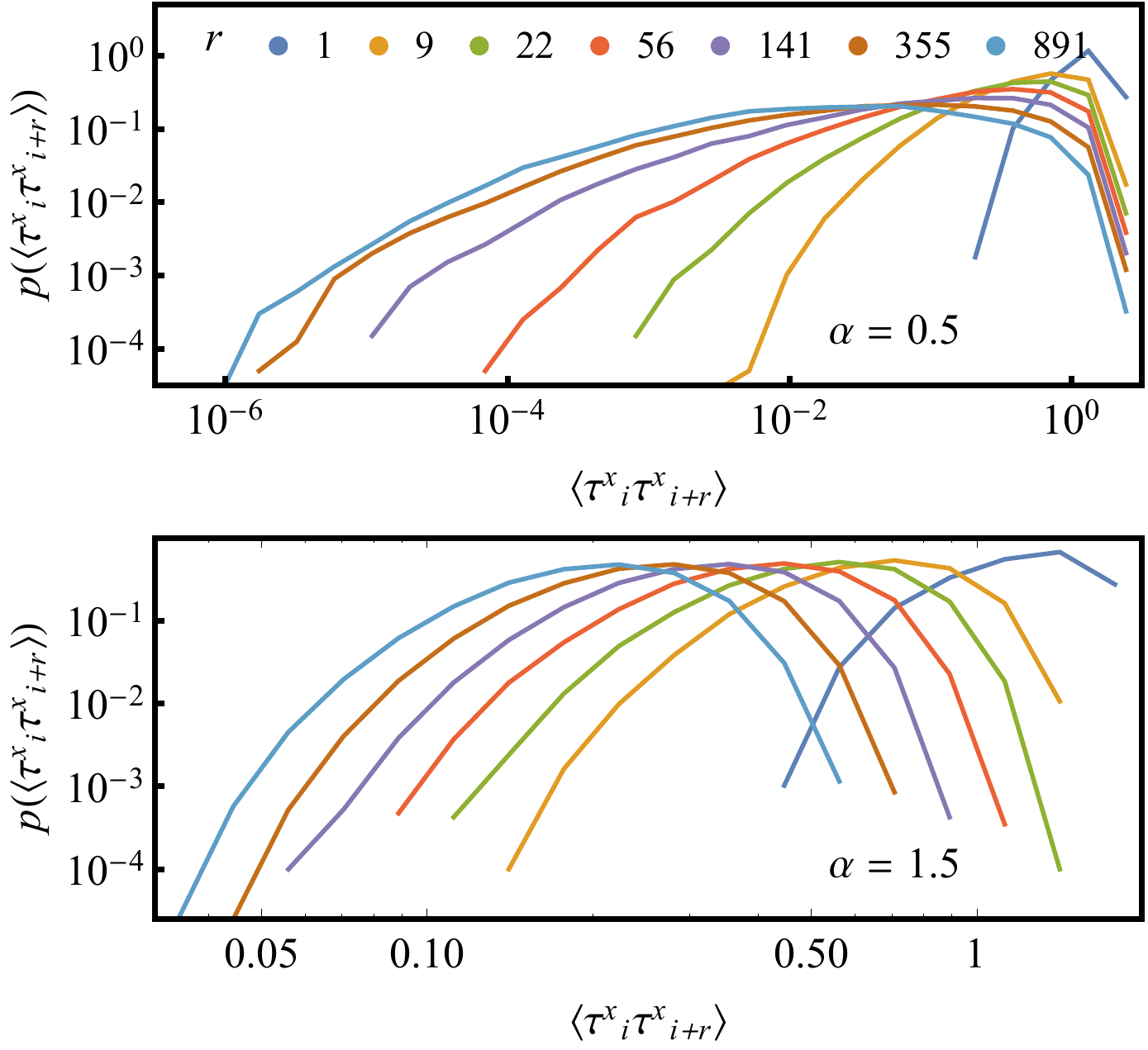}
\caption{Histograms of the order-parameter correlation function $\cexp{\x_i\x_{i+r}}$ for exponentially spaced values of $r$, for weakly hyperuniform (upper) and strongly hyperuniform (lower) systems. In the former case the histograms broaden strongly, while in the latter case they do not broaden. Parameters: same as Fig.~\ref{edcorrs}}
\label{corrhist}
\end{center}
\end{figure}

Our numerical results for the typical and average correlations are plotted in Fig.~\ref{edcorrs}. Both typical and mean correlations behave differently in the two regimes. In the strongly hyperuniform case, we see clean critical behavior in both mean and typical correlations. In the weakly hyperuniform case, typical correlations are consistent with a stretched exponential, with the appropriate exponent $\cexp{\x_i\x_{i+r}} \sim \exp(-\mathrm{const.} r^{\psi})$. Mean correlations clearly decay with a steeper power law than the clean theory would suggest; however, at the accessible system sizes we cannot clearly identify a regime of power-law scaling. A clearer sign of the difference between the two regimes can be seen by considering the histogram of $C_{xx}(r)$ in the two cases (Fig.~\ref{corrhist}). These histograms broaden with $n$ in the weakly hyperuniform case but stay the same width (on a logarithmic scale) in the strongly hyperuniform case, supporting our picture that the weakly hyperuniform case is at infinite randomness. 

\red{
\subsection{Ground state entanglement entropy}
}

\begin{figure}[t]
\begin{center}
\includegraphics[width = 0.4\textwidth]{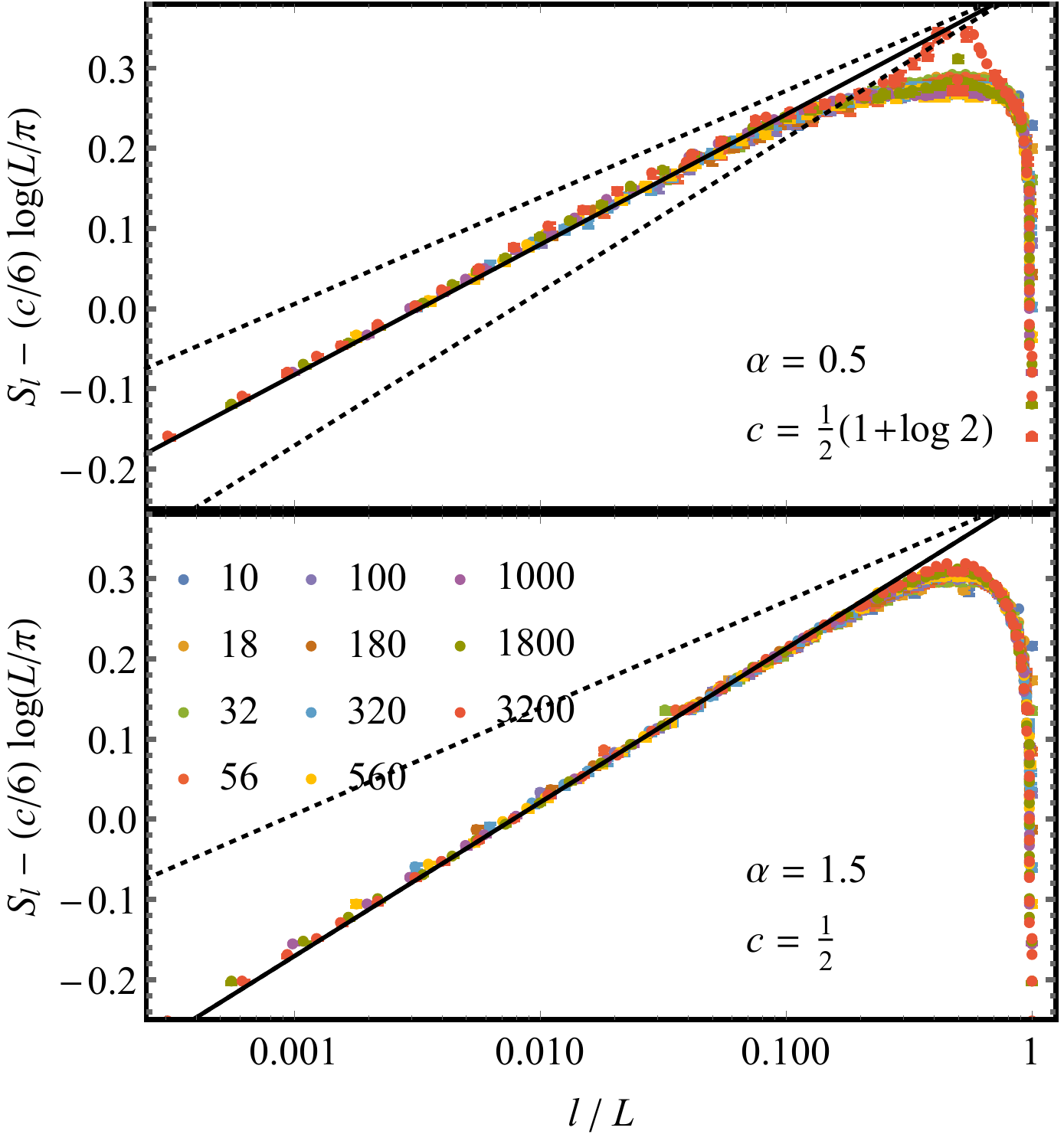}
\caption{
\red{
Entanglement entropy $S_\ell$ of the region ${1 \ldots \ell}$ as a function of $\ell/L$. The predicted growth is $S_\ell \sim (c/6)\log (\ell)$ (solid black line) with the coefficient $c$ set by~\eqref{eq:central_charge}. The collapse of data for different $L$ (legend inset) onto the predicted line for $\ell \ll L$ validates the conjectured form~\eqref{eq:central_charge}. The dashed black lines exhibit the expected growth of the uncorrelated ($\alpha = 0$) and clean Ising ($\alpha = 1$) cases for reference. The panels correspond to different values of $\alpha$ (values inset), with further data given in the Supplemental Material}
}
\label{fig:S}
\end{center}
\end{figure}

\red{
The ground state entanglement entropy provides a useful probe of criticality. In particular, the ground state entanglement $S_\ell$ between the first $\ell$ sites of an open chain and the remaining sites ($\ell +1, \ldots, L$) grows as 
\begin{equation}
    S_\ell = \frac{c}{6} \log \ell + c_1' + \mathrm{O}(\ell/L) 
    \label{eq:SL}
\end{equation}
where the coefficient $c$ forms part of the universal content of the scaling theory, and $c_1'$ is a non universal constant. If the scaling limit is described by a conformal field theory, $c$ is equal to the corresponding central charge. This applies to the clean Ising critical point, where $c = \tfrac12$. In the disordered case there is no underlying conformal field theory, but nonetheless the coefficient $c$ is still fixed by the universal content of the scaling theory. Notably in the case of uncorrelated random disorder, where the transition is described by the infinite randomness critical point, direct calculation yields $c = \tfrac12 \log 2$~\cite{refael2004entanglement}.
}

\red{
In this section we numerically extract the coefficient $c$ from the logarithmic growth of the entanglement entropy~\eqref{eq:SL}. For $\alpha > 1$, we find $c = \tfrac12$, consistent with the clean Ising universality exhibited throughout the strongly hyperuniform regime. In contrast, in the weakly hyperuniform regime, the universality is altered, and we find results which indicate the coefficient $c$ is given by a linear interpolation between the clean Ising value and the uncorrelated infinite randomness value
\begin{equation}
    c = \begin{cases}
    \frac{1}{2} & \text{ for } 1 \leq \alpha \leq 2
    \\
    \frac{\alpha}{2} + (1- \alpha) \frac{\log 2}{2} & \text{ for } 0 \leq \alpha \leq 1.
    \end{cases}
    \label{eq:central_charge}
\end{equation}
}

\red{
In Fig~\ref{fig:S} we plot the numerically extracted form of $S_\ell -  (c/6) \log (L/\pi)$ as a function of $\ell/L$. Plotting in this way causes data corresponding to different $L$ to collapse onto a single scaling form providing $c$ has been correctly identified. In Fig~\ref{fig:S} data is shown for various $L$ (values inset in legend) with the two panels corresponding to values of $\alpha$ from the weakly and strongly hyperuniform regimes respectively. The data collapses, and exhibits good agreement with $(c/6)\log(\ell/L) + c_1'$ (solid black line) where $c$ is set by~\eqref{eq:central_charge}. The dashed black lines exhibit the expected growth $S_\ell$ for the uncorrelated ($\alpha = 0$, $c = \tfrac12 \log 2$) and clean Ising ($\alpha = 1$, $c = \tfrac12$) cases for reference~\footnote{
See Supplemental Material for additional data showing agreement of $S_\ell$ with the form~\eqref{eq:central_charge} over a larger range of values of $\alpha$
}.
}

\section{Discussion}\label{disc}

This work studied a canonical low-dimensional quantum critical point---that of the TFIM---in the presence of random hyperuniform couplings. We have found that the system exhibits two distinct classes of behaviour, dependent on the exponent $\alpha$ (or equivalently the wandering exponent $\beta$). These distinct classes of behaviour are: a line of infinite-randomness critical points, with continuously varying exponent $\psi$; and a ``mixed'' critical point with the equilibrium behavior of the clean Ising model but the dynamics of an insulator. This ``critical Ising insulator'' regime shows that disorder can localize excitations (and thus qualitatively modifying critical dynamics) even when it has minimal effects on equilibrium properties; this is of some general conceptual interest, given that static and dynamical properties are usually intertwined at quantum critical points. Even in the weakly hyperuniform regime, where the strong randomness critical point survives, the Griffiths phases that flank it disappear for hyperuniform couplings, suggesting that hyperuniformity might be a useful knob for controlling Griffiths effects more generally. Using SDRG and perturbation theory, we were able to characterize both critical regimes thoroughly; our predictions are in good agreement with results from exact diagonalization. \red{As is the case with uncorrelated randomness, these results hold for any random modulation with hyperuniform correlations~\eqref{family}, irrespective of other details of the modulation, such as modulation strength or marginal distributions. Generally these additional properties of the modulation play a role only to determine the length-scale above which the asymptotic scaling emerges.}

One might wonder how many distinctively hyperuniform critical phenomena exist beyond the TFIM. 
In general, whenever the control parameter has hyperuniform fluctuations, and there is no source of uncorrelated randomness in the problem, one expects the system will go to a hyperuniform rather than the usual random fixed point. 
Starting from a microscopic model, however, ensuring that the control parameter fluctuations are precisely hyperuniform on all scales might be challenging.
One dimensional models with multiplicative strong-randomness RG rules are a wide class of models where hyperuniformity holds at all scales. 
For such models it is crucial that the couplings on odd and even bonds (or A and B sites) remain separately hyperuniform, as in the TFIM. 
In models such as the TFIM, or spin-1 chains~\cite{hyman_yang}, the odd and even bonds are physically different, so it is natural for them to be separately hyperuniform. In the XXZ chain~\cite{fisher1999phase}, and in toy models of the many-body localization transition~\cite{zbdh, goremykina2018}, this is not the case, and this odd-even structure must be imposed by hand if the model is to flow to a hyperuniform fixed point; otherwise, coarse-graining disrupts the anticorrelations that cause hyperuniformity (Supplemental Material App. D). %
Extending these ideas to more general models, as well as to two-dimensional systems~\cite{motrunich2000infinite}, is an interesting task for future work.

\acknowledgments

We thank Vincenzo Alba, Anushya Chandran, David Huse, and Vadim Oganesyan for helpful discussions. 
The authors acknowledge support from the NSF through Grants No. DMR-1752759 (P.J.D.C.), PHY-1752727 (C.R.L.), and DMR-1653271 (S.G.). P.J.D.C acknowledges the support of the Sloan Foundation. C.R.L. acknowledges support from the Sloan Foundation through a Sloan Research Fellowship. S.G. performed this work in part at the Aspen Center of Physics, which is supported by NSF Grant No. PHY-1607611. S.G. further acknowledges support from a PSC-CUNY internal grant.

\bibliography{QPTFIMcriticality}

\clearpage

\appendix

\section{Broadening of the quasiparticle front}\label{appA}

\begin{figure}[tb]
\begin{center}
\includegraphics[width=0.4\textwidth]{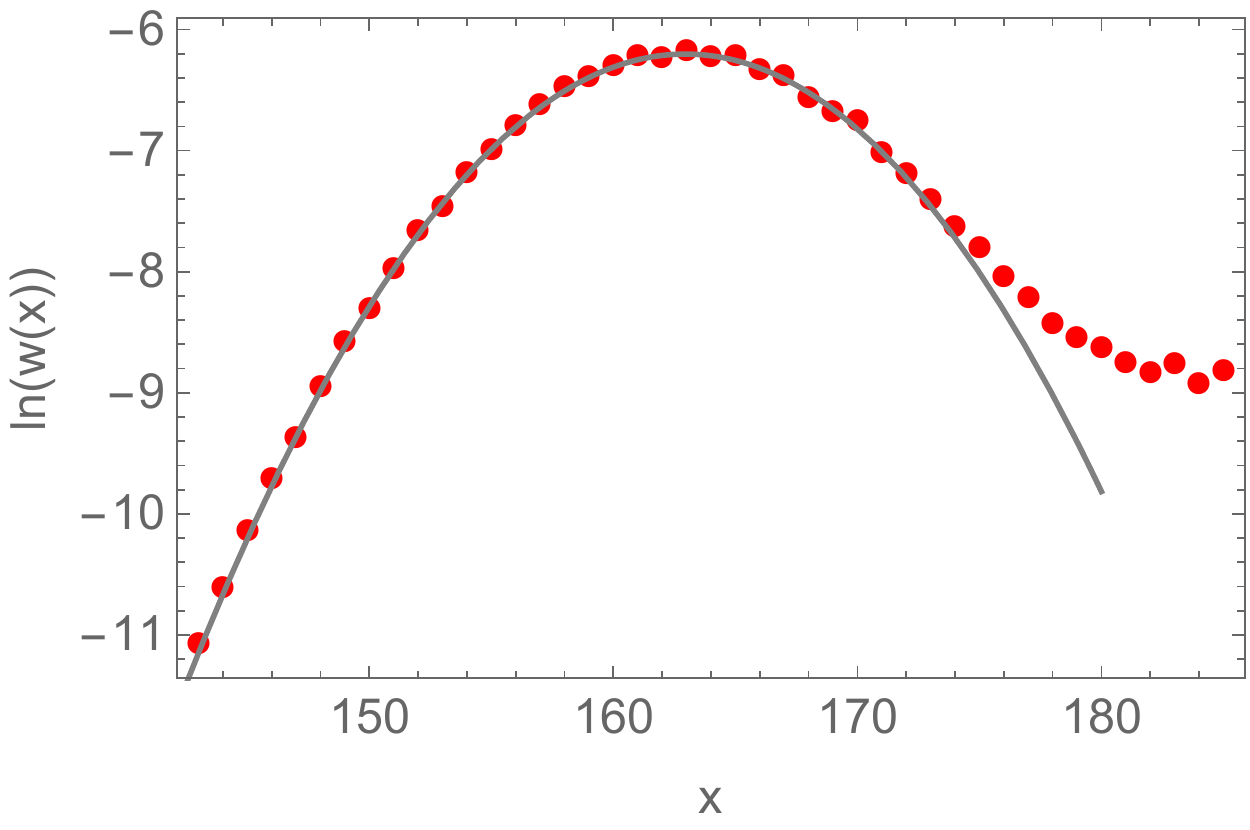}
\includegraphics[width=0.4\textwidth]{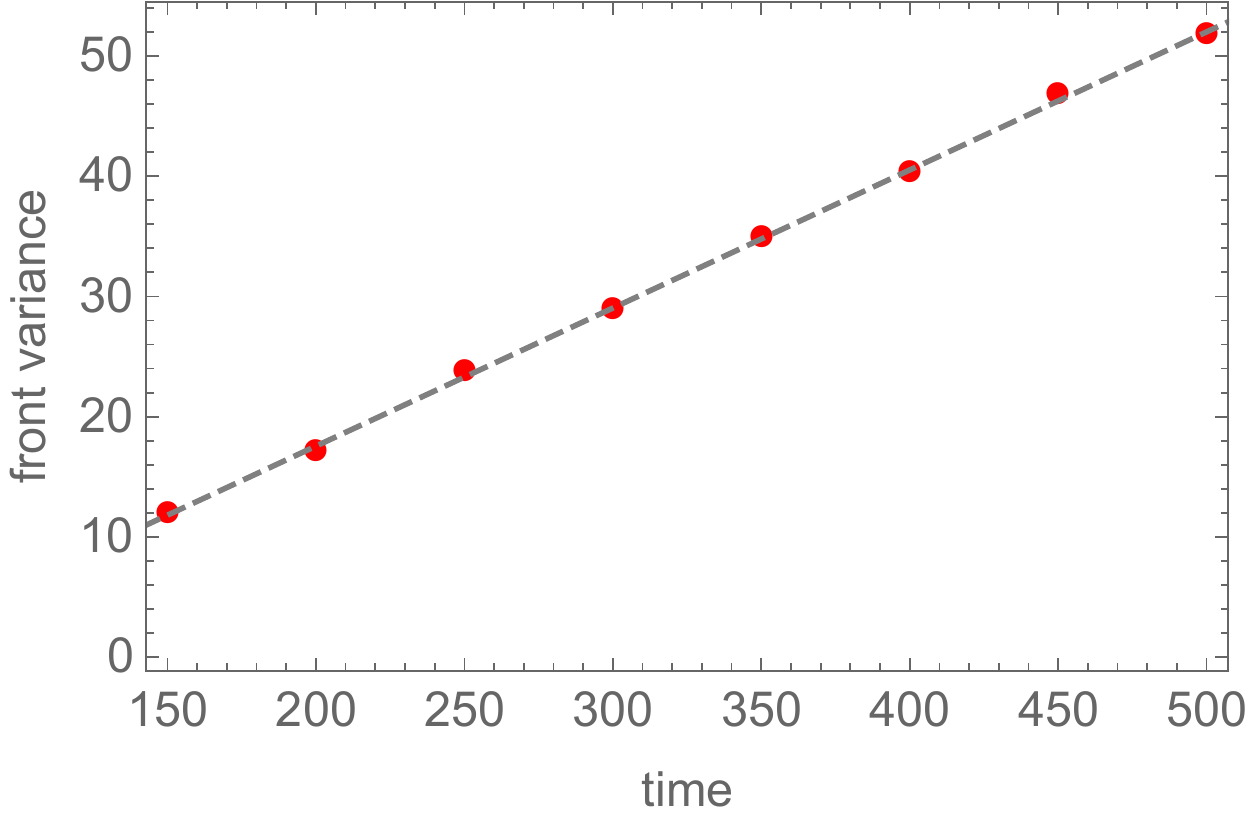}
\caption{Behavior of the front of the wavepacket for $\alpha = 2$. The front is Gaussian in the forward direction (propagating to the left in the upper panel) with variance increasing linearly with time (lower panel).}
\label{frontshapes}
\end{center}
\end{figure}

In this Appendix we present more data on the spreading of wavepackets discussed in Sec.~\ref{wpdyn}. Specifically, Fig.~\ref{frontshapes} presents numerical data on the shape of the wave front at a late time ($t = 400$), showing that its outer tail is Gaussian . 
Inside the wavefront, the shape is not Gaussian, because (as discussed in the main text) a power-law amount of weight is deposited by the wavepacket as it moves. By fitting the outer tail and peak, one can extract a variance from the Gaussian, which we find to grow linearly in time (Fig.~\ref{frontshapes}), in agreement with the analytic predictions in the text. 

\section{Supplementary data from exact diagonalization}
\label{App:ED}

Here we show supplementary data from exact diagonalisation with parameters: $L=2000$ periodic chain, $s=3/16$. (Note that there are $2L$ Majorana modes).

\begin{figure}[tb]
\begin{center}
\includegraphics[width = 0.4\textwidth]{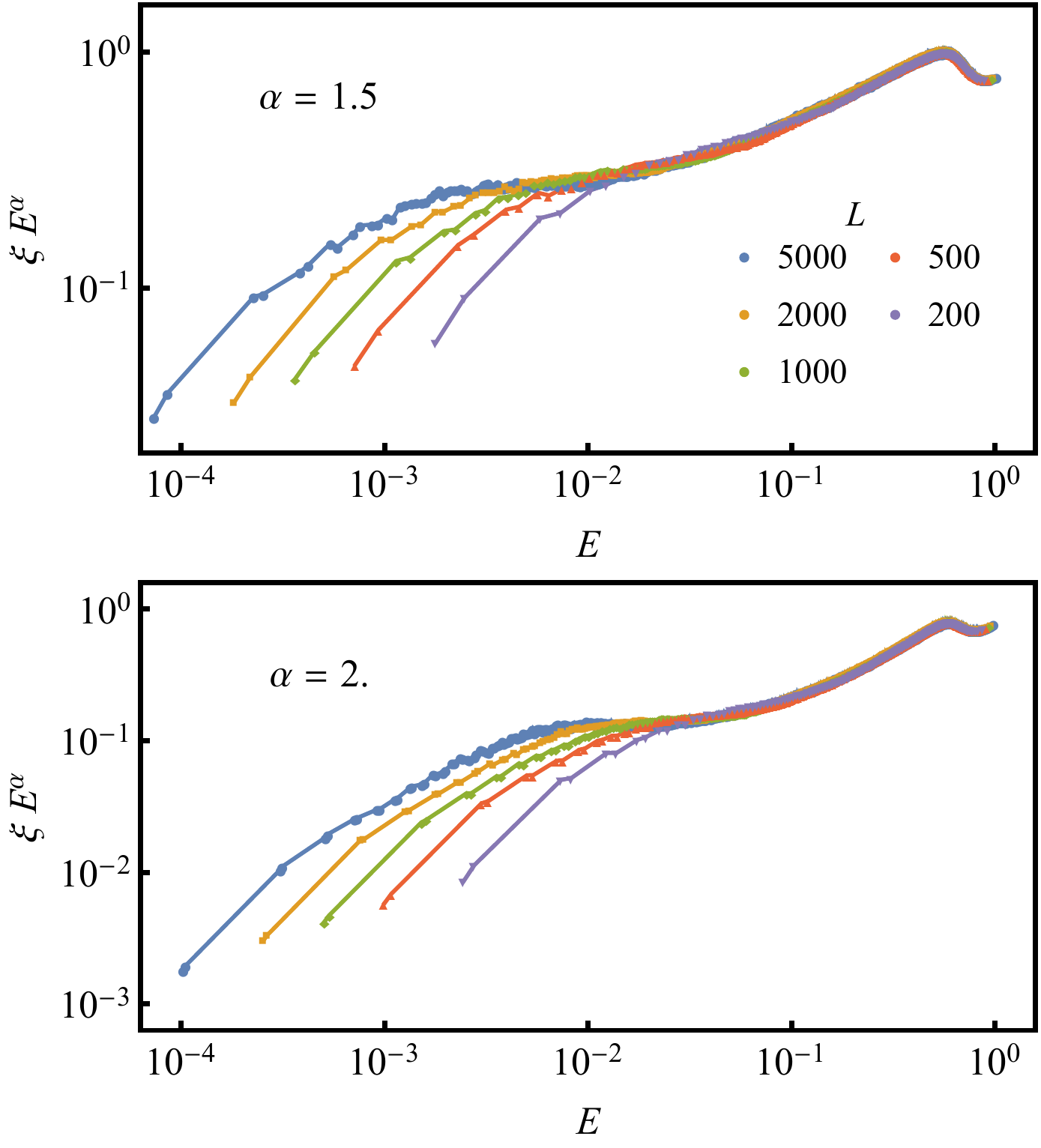}
\caption{
\emph{Low energy behaviour of the localisation length $\xi$ at criticality:}
The predicted low energy behaviour of the localisation length $\xi \sim E^{-\alpha}$ is visible in numerics at small system sizes as the constant valued plateau in $\xi E^\alpha$ whose extent is asymptotically growing with system size $L$.
}
\label{LocLengthL}
\end{center}
\end{figure}

\subsection{Localisation length with strongly hyperuniform disorder}

In Fig.~\ref{LocLengthL} we show numerically evaluated localisation lengths which provide evidence of the $1/\xi \sim E^\alpha$ scaling discussed in Sec.\ref{SH}. This regime is limited, at short distances, by the single-site localized states away from $E = 0$, and at long distances by finite-size effects. However, plotting $\xi E^\alpha$ vs $E$ (Fig.~\ref{LocLengthL}) we see the emergence of the $1/\xi \sim E^\alpha$ dependence, as the level plateau whose extent is growing with increasing system size $L$. 

\subsection{Fermionic correlations}

\begin{figure}[tb]
\begin{center}
\includegraphics[width = 0.4\textwidth]{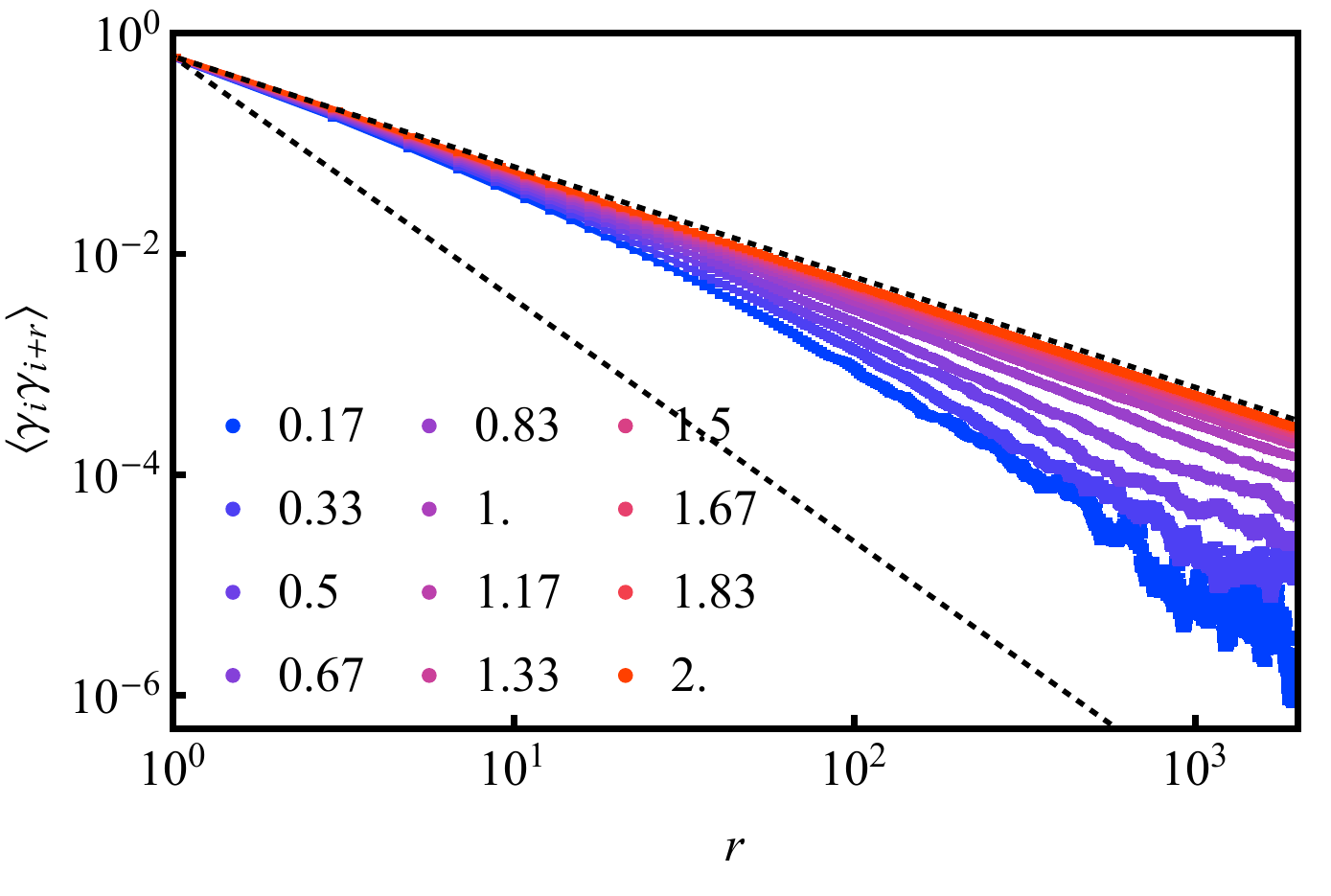}
\includegraphics[width = 0.4\textwidth]{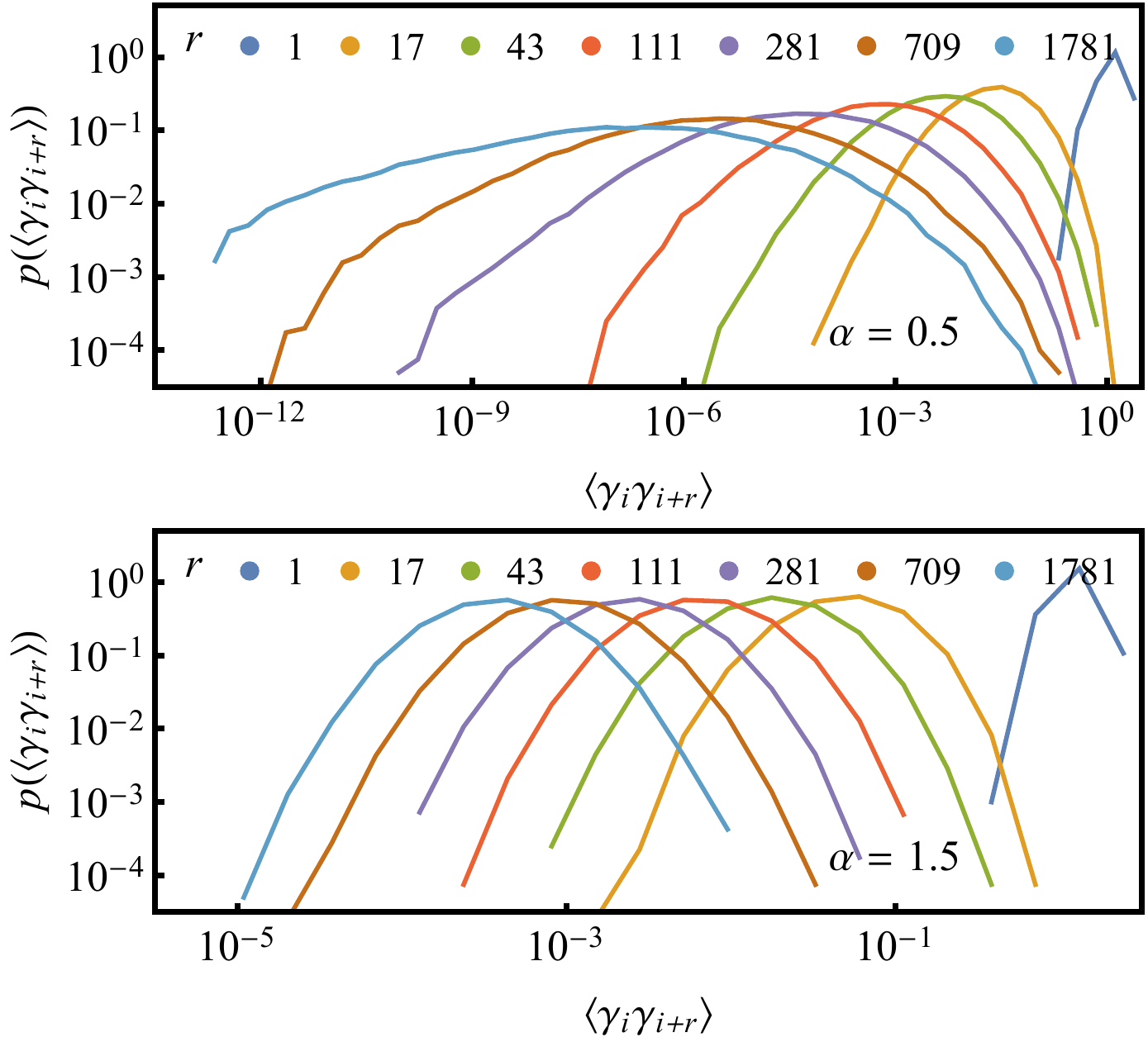}
\caption{
\emph{Fermionic Correlations:} 
Upper panel: $\cexp{\gamma_i\gamma_{i+r}} \sim r^{-2 \Delta_\gamma}$ shown for odd $r$ (correlations are identically zero for even $r$) together with the clean scaling $\Delta_\gamma=1/2$ and iid disorder scaling $\Delta_\gamma=1.1$ (black, dashed lines). 
Lower panels: histograms of disorder for specific exponentially spaced values of $r$. For $\alpha<1$ the correlations becomes increasing broadly distributed at large $r$.
Parameters: $L=2000$, $s=3/16$
}
\label{GamGam}
\end{center}
\end{figure}

In the main text we presented results on the spatial dependence of the equal-time order-parameter correlator. As we show in this Appendix, the results for other equal-time correlators, such as the transverse-field correlator, are qualitatively similar. 
The mean fermionic correlations $\mean{\cexp{\gamma_i\gamma_{i+r}}} \sim r^{-2 \Delta_\gamma}$ are plotted vs $r$ in Fig.~\ref{GamGam} (upper panel) for different values of $\alpha$ (solid lines). The exact diagonalisation suggests the scaling dimension $\Delta_\gamma$ interpolates between the clean value $\Delta_\gamma = 1/2$ (black dashed) and the disordered value $\Delta_\gamma \approx 1.1$~\cite{igloi2018transverse}. 

In Fig.~\ref{GamGam} (lower panels) we see that, as with the spin-correlations (cf. Fig~\ref{edcorrs} and Fig~\ref{corrhist}), the fermionic correlations are broadly distributed for weakly-hyperuniform criticality, leading to a separation between mean and typical correlation, and potentially misleading numerically calculated mean correlations. At strongly hyperuniform criticality, correlations are not broadly distributed and there is no such separation of mean and typical correlations.

\section{Strong disorder RG}
\label{App:SDRG}

The strong disorder renormalisation group (SDRG)~\cite{ma1979random, fisher1992random}, provides an asymptotically exact real space RG on the TFIM with spatially uncorrelated disorder. In this section we outline the relevant details of this RG necessary to show it provides a self-consistent treatment of the weakly hyperuniform models, and furthermore to extract the critical data in these circumstances.

In all numerics in this section we study the symmetry breaking critical point with parameters $L = 10^8$, $\s = 3/16$ and either weakly ($0<\alpha < 1$) or critically ($\alpha=1$) hyper uniform disorder.

\subsection{Recap of SDRG}

At each step in the RG, the strongest energetic scale in the system is identified (i.e. $\max J_i,h_i$) and subsequently decimated, as follows:
\begin{itemize}
\item \emph{Site decimation:} If the strongest scale is a field $h_i$, the spin $\tau_i$ is pinned by this field, decimated from the model. The spins $\tau_{i-1},\tau_{i+1}$ then interact via the ferromagnetic coupling $J'=J_{i}J_{i-1}/h_i$ which is obtained from second order perturbation theory on the pinned spin. 
\item \emph{Bond decimation:} If the strongest scale is the coupling $J_i$, then the spins $\tau_{i},\tau_{i+1}$ are bound into a super spin which is subject to the field $h'=h_ih_{i+1}/J_i$ (similarly obtained from second order perturbation theory), and with a magnetic moment given by the sum of the moments of the bound spins $\mu'=\mu_i+\mu_{i+1}$. 
\end{itemize}
Both of these steps reduce the number of spin degrees of freedom in the system by one. Spins that are merged under the RG are termed a \emph{cluster}.

Following the notation of Fisher~\cite{fisher1992random,fisher1995critical,fisher1999phase}: 
\begin{itemize}
\item The RG cut-off $\Omega$ is set by the largest energy scale in the system $\Omega = \max J_i,h_i$ at each time in the flow.
\item The cut-off is tracked via $\Gamma = \log \Omega_0 - \log \Omega$ where $\Omega_0$ is the initial cut-off. Hence $\Gamma$ is monotonically increasing under the flow.
\item The coupling data is stored via the variables $\beta_i = \log \Omega - \log h_i$, $\zeta_i = \log \Omega - \log J_i$.
\item The number of remaining spins in the system is $N$, with initial value set by the system size $N_0=L$. Thus $l=N_0/N$ records the scale to which the RG has flowed.
\item The mean moment of the active clusters (i.e. those which have not been site decimated) is denoted $\mean\mu$

\end{itemize}

For iid disorder, the flow converges to a fixed point in which these quantities are related by the power laws
\begin{subequations}
\begin{align}
\Gamma \sim \mean\zeta &\sim \mean\beta \sim l^\psi \\
\mean\mu &\sim \Gamma^\phi
\end{align}\label{eq:SDRGscaling}
\end{subequations}
and furthermore the coupling distributions $p_\zeta(\zeta/\mean{\zeta})$, $p_\beta(\beta/\mean{\beta})$ and the moment distribution $p_\mu(\mu/\mean\mu)$ are stable. 
We find corresponding behaviour for weakly hyperuniform disorder with different exponents and scaling distributions $p(\cdot)$. T'he exponents $\psi, \phi$ are related to the decay of the mean and typical correlations
\begin{subequations}
\begin{align}
\mean{\cexp{\tau_i \tau_{i+r}}} & \sim |r|^{-2(1-\psi\phi)} \\
\mean{\log \cexp{\tau_i \tau_{i+r}}} & \sim -|r|^\psi.
\end{align}
\end{subequations}
In the iid case, these take the familiar values $\psi=1/2$, $\phi = (1 + \sqrt{5})/2$.

\subsection{Asymptotic exactness for weakly hyperuniform disorder}

As the SDRG flows, physical spins combine into clusters which are eventually decimated from the system. 
In principle, treating the decimation to second order in perturbation theory introduces errors. 
However, the error in each individual decimation is small when the decimated effective coupling $\Omega$ is much larger than the neighbouring couplings $h_i,J_i$, or equivalently when $\zeta_i,\beta_i \gg 0$. 

In the iid case, there are no disorder correlations at any stage in the flow and the neighbouring log couplings are characterised by the corresponding global mean values $\mean\zeta, \mean\beta$. The RG flow is asymptotically exact if $\mean\zeta, \mean\beta$ increase without bound under the flow, since under this assumption the decimations at late stages of the RG are increasingly accurate.

In the hyperuniform case the disorder is anticorrelated (disorder fluctuations are suppressed on all length scales, see Sec.~\ref{model}). The couplings in the vicinity of the maximal coupling $\Omega$ are thus typically smaller than the mean log couplings $\zeta_i \gtrsim \mean\zeta$, $\beta_i \gtrsim \mean\beta$. The condition that $\mean\zeta, \mean\beta$ grow without bound presumably remains sufficient for the asymptotic exactness of the SDRG, since these local anticorrelations further suppress decimation errors relative to the iid case. 

The asymptotic growth of $\mean\zeta, \mean\beta$ for weakly hyperuniform disorder $\alpha < 1$ is seen in Fig.~\ref{rsrgresults} (upper panel).

For strongly hyperuniform disorder $\alpha >1$ one finds that $\mean\beta,\mean\zeta$ are asymptotically decreasing with $l$, and hence the SDRG introduces significant error at every stage in the flow and cannot be applied.

In the critical case $\alpha = 1$ $\mean\zeta, \mean\beta$ appear to increase slowly over numerically accessible dynamical ranges, and so we include this case in our SDRG analysis.

\subsection{Convergence to strong disorder fixed point}

\begin{figure}[tb]
\begin{center}
\includegraphics[width = 0.4\textwidth]{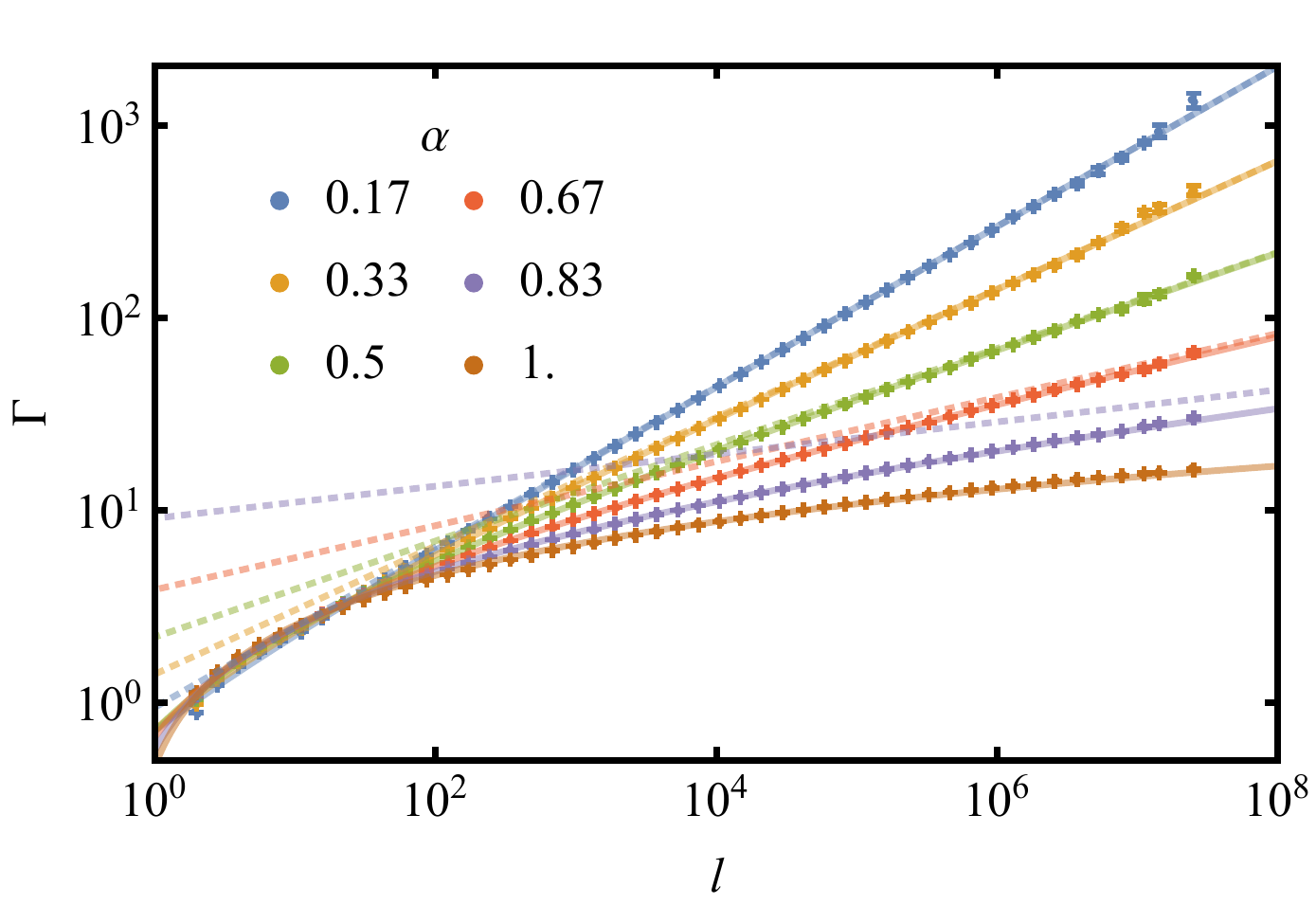}
\caption{
\emph{Energy cut-off scaling:}
The relationship $\Gamma \sim l^\psi$ is asymptotically observed. Plot points show SDRG data. Error bars show standard error on the mean. Solid line shows a non-linear fit of $\Gamma = a l^{(1-\alpha)/2} - b$ for fit parameters $a,b$. Dashed line shows the $\Gamma = a l^{(1-\alpha)/2}$ for the same value of $a$.
}
\label{SDRGgamma}
\end{center}
\end{figure}

\begin{figure}[tb]
\begin{center}
\includegraphics[width = 0.4\textwidth]{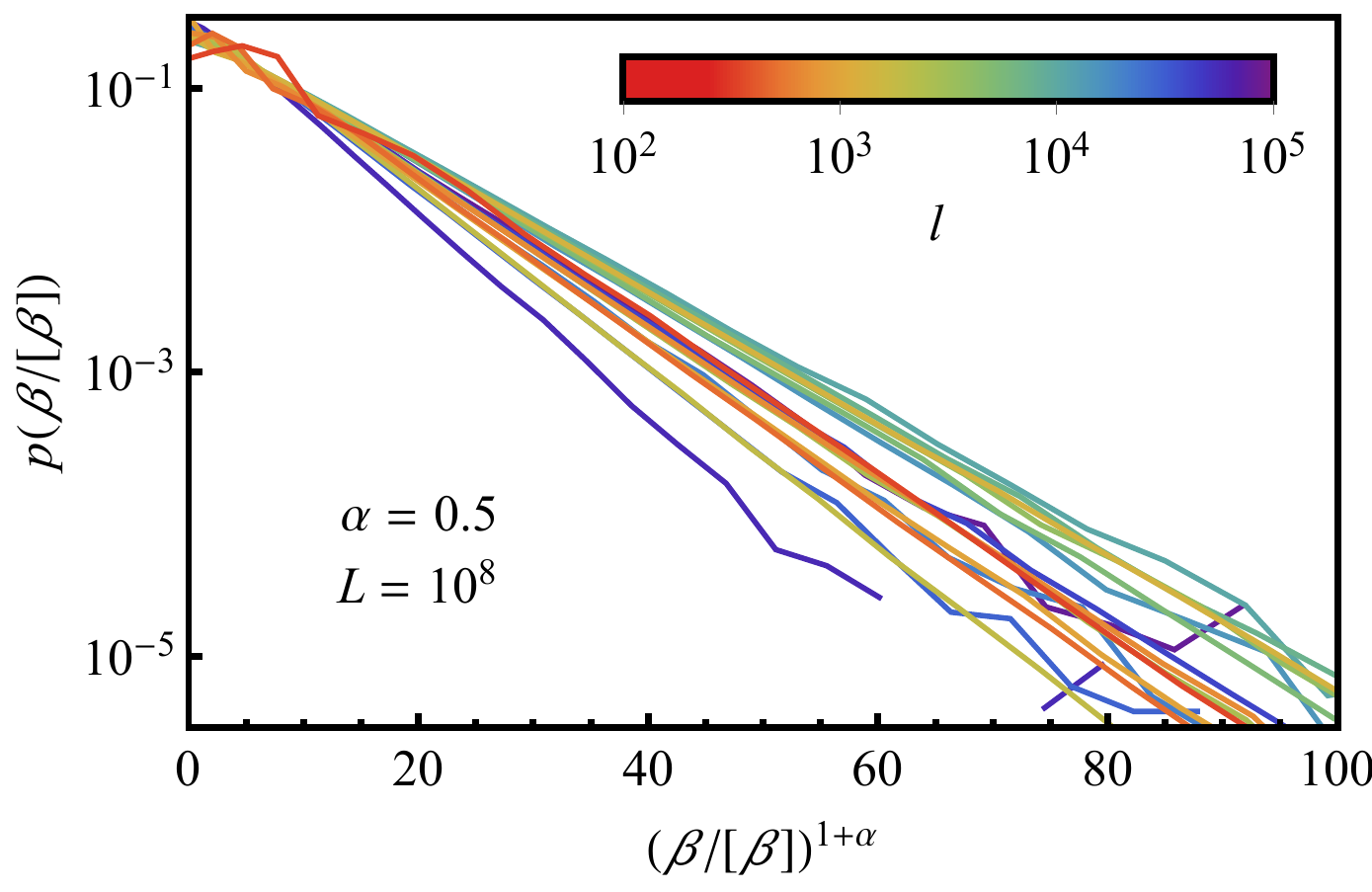}
\includegraphics[width = 0.4\textwidth]{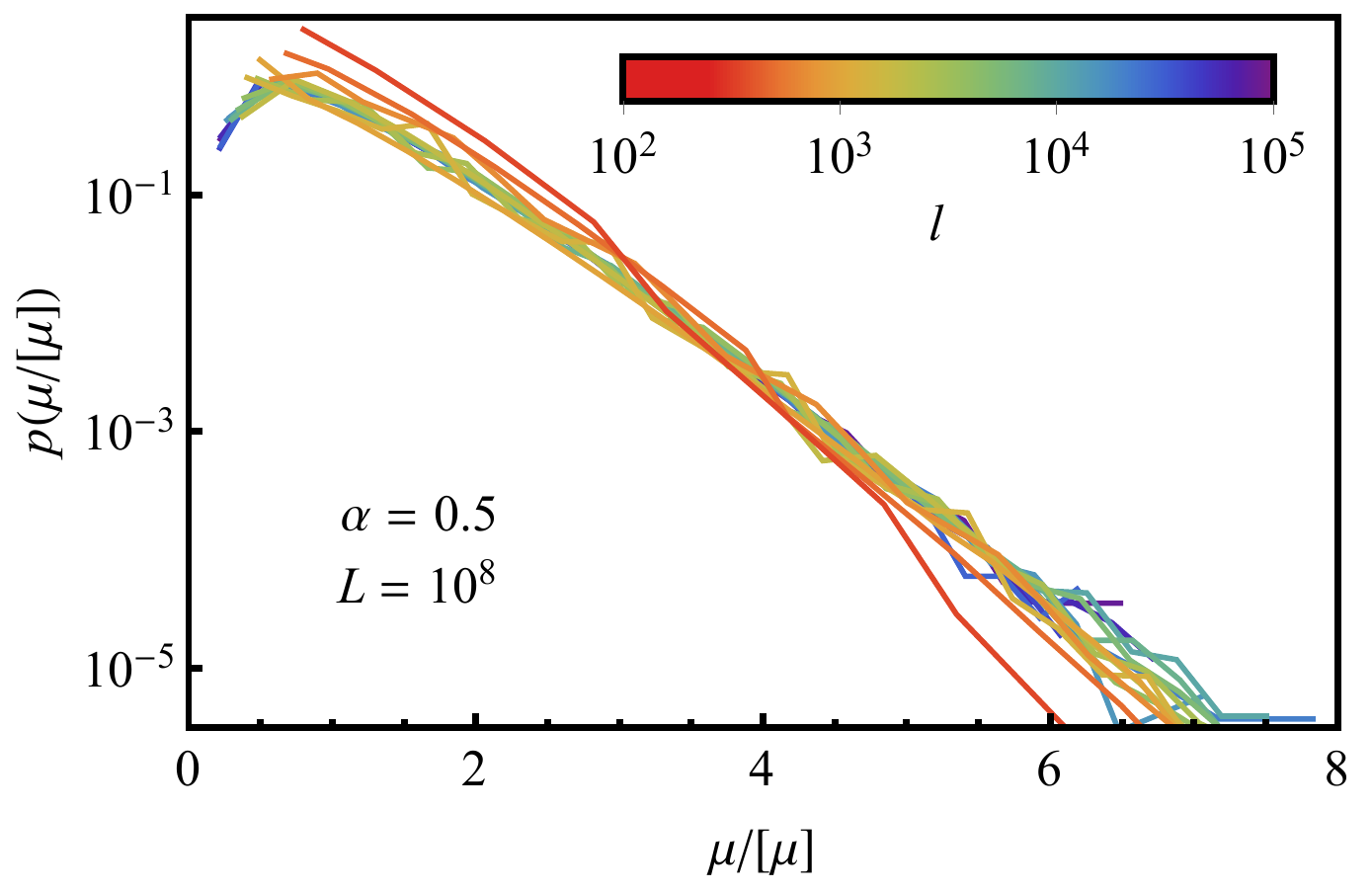}
\caption{
\emph{Stability of distributions within a single SDRG realisation:}
Upper panel: $p_\beta(\beta/\mean\beta)$ vs $(\beta/\mean\beta)^{1+\alpha}$ is plotted for $20$ log-spaced values of $l$. Intra-sample fluctuations lead to variation in the horizontal scaling which grows as the number of remaining couplings becomes small (at large $l$). Otherwise the shape of the distribution remains stable un the SDRG flow. Lower Panel: $p_\mu(\mu/\mean\mu)$ vs $\mu/\mean\mu$ is plotted for the same SDRG realisation. Discrepancies at small $l$ decrease with increasing $l$ as the RG approaches the fixed point behaviour. Both plots are for $\alpha = 0.5$.
}
\label{SDRGDistStab}
\end{center}
\end{figure}

The convergence of the flow is seen by the stability of the relationships~\eqref{eq:SDRGscaling}, the coupling distributions $p_\zeta(\zeta/\mean{\zeta}), p_\beta(\beta/\mean{\beta})$ and the moment distribution $p_\mu(\mu/\mean\mu)$.

As the model is self-dual under Ising symmetry, necessarily $\mean{\zeta} = \mean{\beta}$. Averaging these quantities together the relationship $\mean{\zeta} = \mean{\beta} \sim l^\psi$ is seen to be stable in Fig.~\ref{rsrgresults} (upper panel).

The asymptotic relationship $\Gamma \sim l^\psi$ is seen in Fig.~\ref{SDRGgamma} for weakly hyper uniform disorder: here data from SDRG flows is plotted, this data shows agreement to the fit $\Gamma = a l^\psi - b$ (solid lines) for the same values of $\psi$ as Fig.~\ref{rsrgresults}, and asymptotically converges to the power law $\Gamma = a l^\psi$ (dashed lines). The critical case $\alpha=1$ shows agreement to the behaviour $\Gamma \sim \log l$ (solid brown line).

The stability of the coupling distributions $p_\beta(\beta/\mean{\beta})$ (and by Ising symmetry the distribution $p_\zeta(\zeta/\mean{\zeta})$) and the moment distribution $p_\mu(\mu/\mean\mu)$ is seen in Fig.~\ref{SDRGDistStab}. In this figure histograms of the data are plotted for snapshots of the SDRG at different scales $l$ within a single realisation of the SDRG flow. For $p_\zeta$, $p_\beta$ there are fluctuations within the  sample that lead to discrepancies between the distributions as the number of the remaining couplings becomes small (at large $l$). For $p_\mu$ the variation is smaller and more significant at small $l$, after which the flow behaviour converges to the fixed point behaviour. In both cases the overall behaviour indicates the stability of the RG fixed point.

\subsection{SDRG fixed point properties}

\subsubsection{Exponent $\psi$}

The SDRG shows numerically that $\psi = (1-\alpha)/2$. As is noted in the main text this result can be seen intuitively: the energy cut-off $\Omega$ is the largest renormalised coupling and hence has the form 
\begin{equation}
\Omega = \frac{J_i J_{i+1} \cdots J_{i+r}}{h_i h_{i+1} \cdots h_{i+r-1}}
\end{equation}
(or equivalently with $J,h$ interchanged), where $r=O(l)$. For hyperuniform disorder (Sec.~\ref{model}) it follows that
\begin{equation}
\Gamma \sim \log \Omega \sim \sum_{j=i}^{i+l-1} q_i \sim
\begin{cases}
l^{(1-\alpha)/2} & \text{ for } \alpha < 1, \\
\log l & \text{ for } \alpha = 1, \\
1 & \text{ for } \alpha > 1.\\
\end{cases}
\end{equation}
The scaling $\Gamma \sim l^{(1-\alpha)/2}$ is numerically verified for $\alpha < 1$ in Fig.~\ref{SDRGgamma}: here data from SDRG flows is plotted with error bars showing the standard error on the mean from $3$ SDRG realisations. In Fig.~\ref{SDRGgamma} we see agreement to the fit $\Gamma = a l^{(1-\alpha)/2} - b$ (solid lines) and asymptotically convergence to the power law $\Gamma = a l^{(1-\alpha)/2}$ (dashed lines). The critical case $\alpha=1$ shows agreement to the behaviour $\Gamma = a \log l + b$ (solid brown line) where $a$ is found to depend on the details of the disorder distribution, in this case exhibiting a linear dependence on the disorder strength $a \sim s$.

Similarly the exponent values $\psi = (1-\alpha)/2$ are used in the fits (solid lines) in Fig.~\ref{rsrgresults} (upper panel), and show good agreement to the data, indicating that expected $\mean{\zeta},\mean{\beta} \sim l^{(1-\alpha)/2}$

The values of $\psi$ extracted from fitting with to $\Gamma = a l^{\psi} - b$ with $a,b,\psi$ as free parameters are compared with the exact form $\psi = (1-\alpha)/2$ are compared in Fig.~\ref{whc} (upper panel) and show good agreement.

\subsubsection{Exponent $\psi\phi$}
\label{sec:PP}

\begin{figure}[tb]
\begin{center}
\includegraphics[width = 0.4\textwidth]{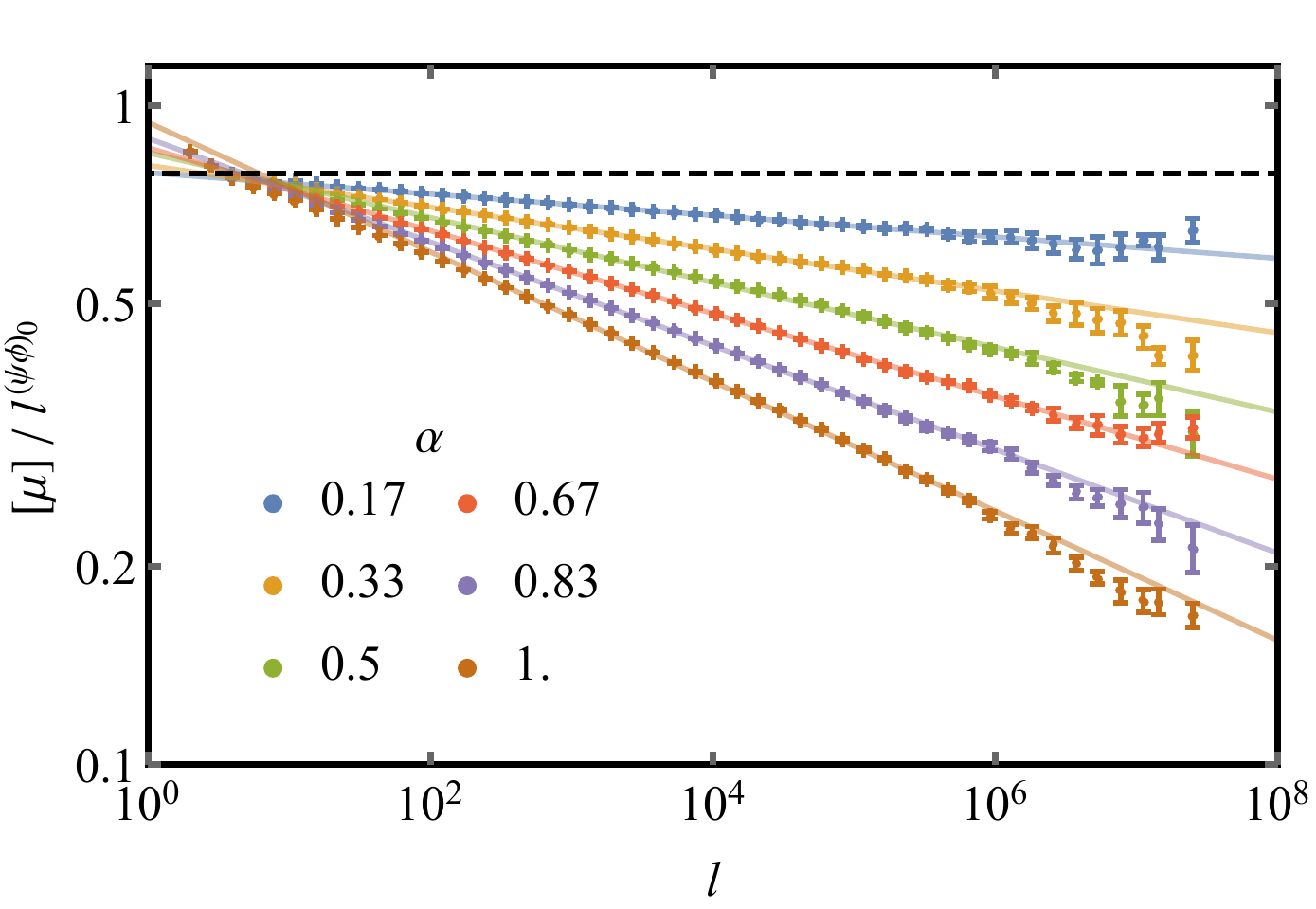}
\caption{
\emph{Scaling of mean moment:} The mean moment scales as $\mean\mu \sim l^{\psi\phi}$, we rescale by $l^{(\psi\phi)_0}$ (with $(\psi\phi)_0 = (1+\sqrt{5})/4$, the iid value), and here plot $\mean\mu/l^{(\psi\phi)_0}$ versus $l$ (coloured points), standard error on the mean for each point is shown. Linear fits are shown (coloured dashed) from which exponent estimates are extracted. For comparison the iid scaling is shown (black dashed).
}
\label{SDRGmu}
\end{center}
\end{figure}

The power law relationship $\mean\mu \sim l^{\psi\phi}$ (see eq.~\eqref{eq:SDRGscaling}) is verified for $\alpha\leq 1$ in Fig.~\ref{SDRGmu}. The extracted values of the exponent $\psi\phi$ are shown in Fig.~\ref{whc} (lower panel, blue triangles), where one sees that $\psi\phi$ exhibits a simple linear dependence of $\alpha$.

\subsubsection{Exponent $\Delta_\sigma$}
\label{sec:DS}

The power law decay of the mean critical spin-spin correlations $\mean{\cexp{\x_i\x_j}}$ can be accessed directly from the SDRG.

As the SDRG flows, spins are decimated into clusters of strongly correlated spins, whilst correlations between these clusters are determined by the remaining reduced couplings $\beta_i,\zeta_i$. The asymptotic growth of $\mean\beta,\mean\zeta$ means that inter-cluster correlations are asymptotically weaker than intra-cluster correlations.

This implies that spins within a cluster are close to maximally correlated whilst correlations between spins from different clusters are asymptotically weaker.
Thus, we write
\begin{equation}
\cexp{\x_i\x_j} =
\begin{cases}
1 & \text{ if sites $i$, $j$ in the same cluster} \\
0 & \text{ otherwise}. \\
\end{cases}
\end{equation}
Calculating $\cexp{\x_i\x_{i+r}}$ in this way, we then average over $i$, and disorder to obtain the correlations shown in Fig.~\ref{rsrgresults} (lower panel). The extracted values of the spin scaling dimension $\Delta_\sigma$ (defined via $\mean{\cexp{\x_i\x_{i+r}}} \sim r^{-2 \Delta_\sigma}$) are shown in Fig.~\ref{whc} (lower panel, yellow triangles). 

\subsubsection{Relationship $\Delta_\sigma = 1 - \psi\phi$}
\label{sec:DelPP}

\begin{figure}[tb]
\begin{center}
\includegraphics[width = 0.4\textwidth]{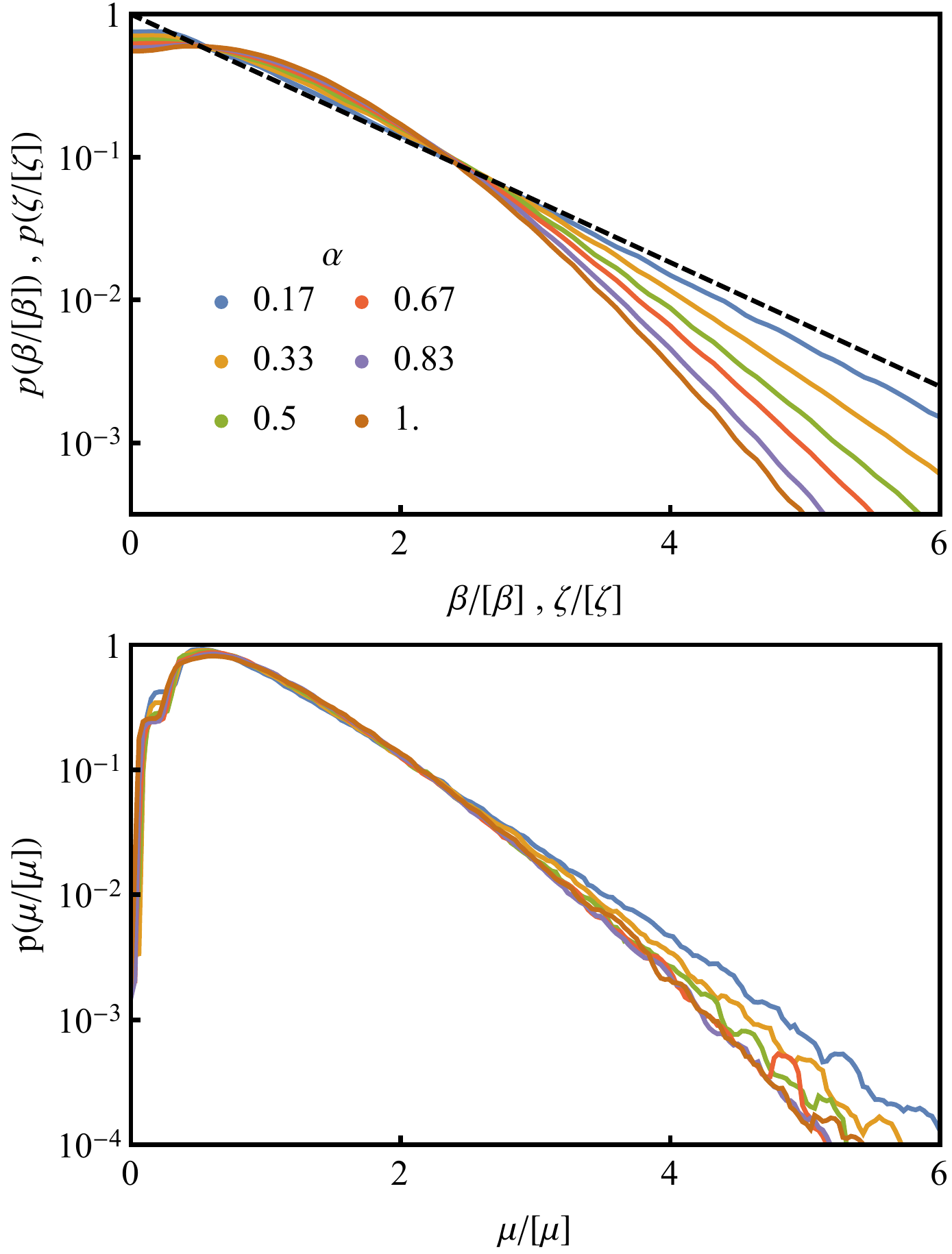}
\caption{
\emph{SDRG fixed point distributions:}
Upper panel: For iid disorder the coupling distribution is given by an exponential (black dashed) numerically obtained distributions are shown obtained for weakly hyper uniform modulation (solid colours). Lower panel: The moment distributions are shown for the same values of $\alpha$.
}
\label{SDRGdist}
\end{center}
\end{figure}

The scaling of $\mean\mu$ can be estimated the spin scaling dimension. Consider running the RG until the system is rescaled by a factor $l \gg 1$. The probability that a given physical spin is still part of an active cluster (i.e. has not been site decimated) is given by $\mean\mu/l = l^{\psi\phi-1}$. For two spins to be strongly correlated they must be part of the same cluster. For two spins to be bond decimated into the same cluster, they must first become neighbours under the RG flow. The probability that two spins initially separated by a length $r$ both survive, to become neighbours under the flow is hence $(\mean{\mu}/l)^2_{l=r} = (r^{\psi\phi-1})^2 = r^{-2(1-\psi\phi)}$. This argument implies that $\mean{\cexp{\x_i\x_{i+r}}} = (\mean{\mu}/l)^2_{l=r}$ and hence that $\mean{\cexp{\x_i\x_{i+r}}} \sim r^{-2(1-\psi\phi)}$ and thus $\Delta_\sigma = 1-\psi\phi$.

This relation is confirmed numerically in Fig.~\ref{whc}. 
The extracted values of $\Delta_\sigma$ (Sec.~\ref{sec:DS}, yellow triangles), show good agreement with the values $1 - \psi \phi$ (Sec.~\ref{sec:PP}, blue triangles).

\subsubsection{Coupling and moment distributions}

\begin{figure*}
    \centering
    \includegraphics[width = \textwidth]{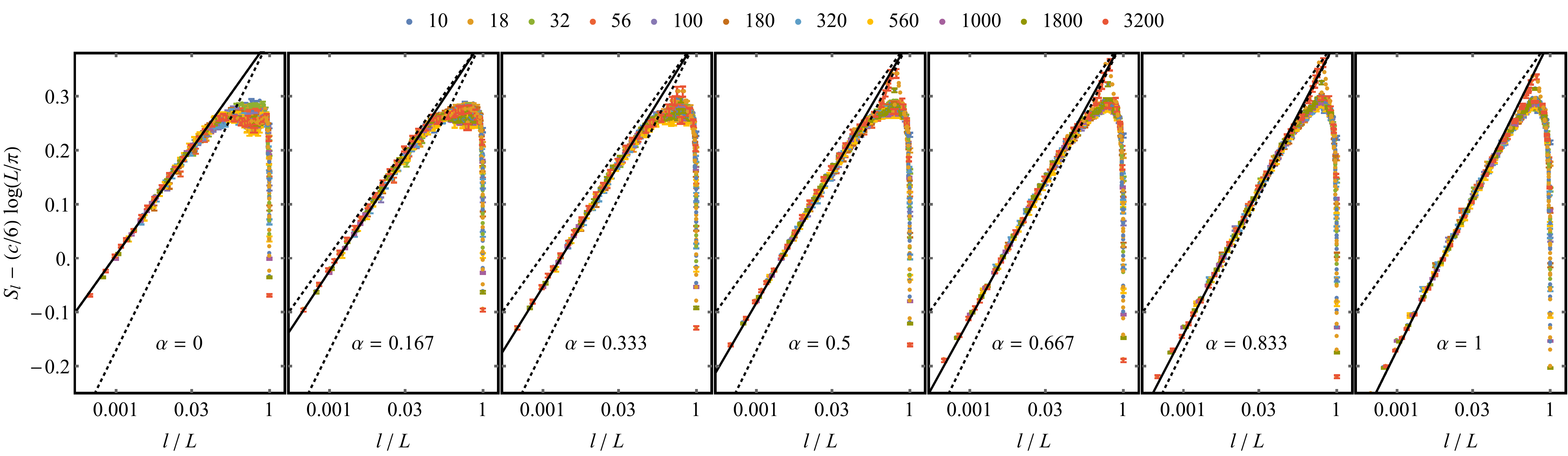}
    \caption{\red{Entanglement entropy $S_\ell$ of the region ${1 \ldots \ell}$ as a function of $\ell/L$. $S_\ell$ is given by Eq.~\eqref{eq:SL} with the coefficient $c$ set by Eq.~\eqref{eq:central_charge}. The collapse of data for different $L$ (legend, top) onto the predicted line (black solid) for $\ell \ll L$ validates the conjectured form~\eqref{eq:central_charge}. The dashed black lines exhibit the expected growth of the uncorrelated ($\alpha = 0$) and clean Ising ($\alpha = 1$) cases for reference. Each panel corresponds to different value of $\alpha$ (values inset), taken from the weakly hyperuniform regime.}}
    \label{fig:S_supp}
\end{figure*}

In the iid case, the SDRG fixed point couplings follow the distributions $p_\zeta(\zeta/\mean{\zeta}), p_\beta(\beta/\mean{\beta})$ where $p_\zeta(x) = p_\beta(x) = \mathrm{e}^{-x}$. 
In the the weakly hyperuniform case, we find numerically that distributions are narrower, with asymptotic tails
\begin{equation}
\log p_\zeta(x) = \log p_\beta(x) \sim -x^{1+\alpha}, 
\label{eq:distscaling}
\end{equation}
and furthermore are peaked at $x>0$. The decreased weight at large and small $x$ indicates that both very large and very small couplings are rarer than in the iid case. 

The coupling distributions found at the RG fixed points are shown in Fig.~\ref{SDRGdist} (upper panel). The distribution, and hence the asymptotic form~\eqref{eq:distscaling}, is seen to be stable under the SDRG flow in Fig.~\ref{SDRGDistStab}. One similarly finds that the moment distribution is similarly altered, with asymptotically less weight in the tails Fig.~\ref{SDRGdist} (lower panel) though this effect is less significant.

\red{
\section{Ground state entanglement entropy: additional data}
}

\red{
In this section we present further numerical study of the dependence of bipartite entanglement entropy coefficient $c$ in Eq.~\eqref{eq:SL} on the hyperuniformity parameter $\alpha$. In the main text we conjectured the form Eq.~\eqref{eq:central_charge}, and we here present further supporting numerical evidence. In the strongly hyperuniform regime, $c$ is conjectured to be unaltered from the value of the clean Ising central charge, this is shown for $\alpha = 1.5$ in Fig~\ref{fig:S}, and verified for other $1 < \alpha < 2$ (data not shown). The unaltered value of the coefficient $c$ in the presence of strongly hyperuniform modulation is consistent with the clean Ising universality exhibited throughout this regime. In contrast in the weakly hyperuniform regime, we conjectured that the coefficient $c$ (Eq.~\eqref{eq:central_charge}) is given by
\begin{equation}
    c = \frac{\alpha}{2} + (1- \alpha) \frac{\log 2}{2}, \quad \text{for} \quad 0 < \alpha < 1
    \label{eq:c_supp}
\end{equation}
i.e. by a linear interpolation between its values at the clean Ising critical point, and at the infinite randomness fixed point with uncorrelated disorder ($\alpha = 0$).
}

\red{
In Fig~\ref{fig:S_supp} we show data analogous to Fig~\ref{fig:S} for values of $\alpha$ (inset) taken throughout the weakly hyperuniform regime. As in the figures in the main text, each panel shows collapse of $S_\ell - (c/6) \log (L / \pi)$ (with $c$ given by~\eqref{eq:c_supp}) as a function of the rescaled cut position $\ell/L$. Data is shown for various $L$ (values in legend, above Fig~\ref{fig:S_supp}). The collapse exhibited in each plot, and the agreement with the proposed theoretical form (solid black line) supports the conjectured form Eq.~\eqref{eq:central_charge}. In each plot the dashed lines indicate the scaling obtained for the uncorrelated ($\alpha = 0$, $c = \tfrac12 \log 2$) and clean Ising ($\alpha = 1$, $c = \tfrac12$) cases for reference.
}

\section{Importance of bipartite structure}\label{bipartite}

\begin{figure}[t]
\begin{center}
\includegraphics[width = 0.4\textwidth]{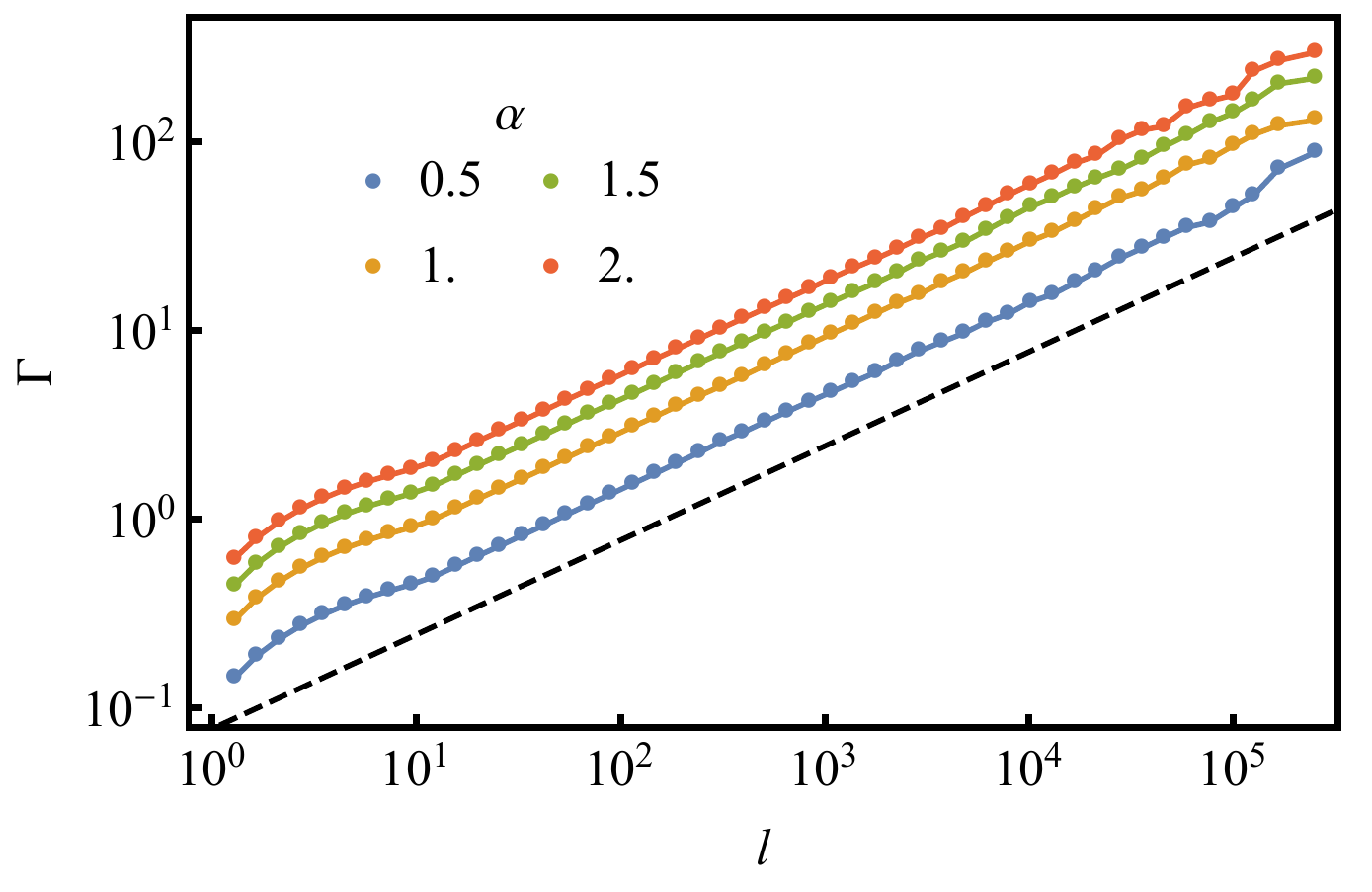}
\caption{
Wandering of effective couplings $\log(J^{\mathrm{even}}_i / J^{\mathrm{odd}}_i)$ where only the full sequence is hyperuniform (as in the XXZ antiferromagnet). The reduced couplings $\delta_i = \log(J^{\mathrm{even}}_i / J^{\mathrm{odd}}_i)$ are no longer hyperuniform. Data is shown for different values of $\alpha$ (solid colors) in a single RG run, all runs exhibit $\Gamma \sim l^\psi$ with $\psi = 1/2$ (black dashed).
For comparison, the case where each subsequence is separately hyperuniform (as in the Ising model) is in Fig.~\ref{SDRGgamma}. 
Parameters: $L=10^6, s=3/16$
}
\label{figbipartite}
\end{center}
\end{figure}

As noted in the main text, a crucial feature of the SDRG rules for the TFIM is that they preserve the bipartite structure of the problem: an effective bond, at any stage of the RG, is a product of $J$'s divided by a product of $h$'s, and vice versa for an effective transverse field. This feature allowed us to argue (Sec.~\ref{griffrg}) that the wandering behavior of the control parameter $\log (J_i/h_i) $---though possibly not each separately---stays unchanged after any finite number of RG steps. In the TFIM this structure emerges naturally from the microscopic problem; however, in models such as the XX chain it is more natural to have all the bonds drawn from the same hyperuniform distribution. Such a pattern will not stay hyperuniform under SDRG. Generically, subsequences (e.g., all the even elements) of a hyperuniform sequence are not themselves hyperuniform, since picking out subsequences spoils the cancellations that give rise to hyperuniformity. The SDRG preserves only the hyperuniformity of the even sites or the odd sites; if these subsequences are not themselves hyperuniform, the distribution of effective couplings after SDRG ceases to be hyperuniform, and, indeed, looks essentially uncorrelated; see Fig.~\ref{figbipartite}.

\end{document}